\documentclass[11pt,a4paper]{article}
\usepackage{jheppub}

\begin{document}

\title{Noncommutative U(1) gauge theory from a worldline perspective}
\author[1]{Naser Ahmadiniaz,}
\author[1,2]{Olindo Corradini.}
\author[3]{Daniela D'Ascanio,}
\author[1]{Sendic Estrada-Jim\'enez}
\author[3]{and Pablo Pisani}

\affiliation[1]{Facultad de Ciencias en F\'isica y Matem\'aticas,
Universidad Aut\'onoma de Chiapas,\\ Ciudad Universitaria, Tuxtla Guti\'errez 29050, M\'exico}
\affiliation[2]{Dipartimento di Scienze Fisiche, Informatiche e Matematiche,\\ Universit\`a di Modena e Reggio Emilia,\\ Via Campi 213/A, I-41125 Modena, Italy}
\affiliation[3]{Instituto de F\'isica La Plata, CONICET -- Universidad Nacional de La Plata,\\ CC 67 (1900), La Plata, Argentina}

\emailAdd{naser@ifm.umich.mx}
\emailAdd{olindo.corradini@unach.mx}
\emailAdd{dascanio@fisica.unlp.edu.ar}
\emailAdd{sestrada@unach.mx}
\emailAdd{pisani@fisica.unlp.edu.ar}

\abstract{We study pure noncommutative $U(1)$ gauge theory representing its one-loop effective action in terms of a phase space worldline path integral. We write the quadratic action using the background field method to keep explicit gauge invariance, and then employ the worldline formalism to write the one-loop effective action, singling out UV-divergent parts and finite (planar and non-planar) parts, and study renormalization properties of the theory. This amounts to employ worldline Feynman rules for the phase space path integral, that nicely incorporate the Fadeev-Popov ghost contribution and efficiently separate planar and non-planar contributions. We also show that the effective action calculation is independent of the choice of the worldline Green's function, that corresponds to a particular way of factoring out a particle zero-mode. This allows to employ homogeneous string-inspired Feynman rules that greatly simplify the computation.}

\maketitle

\section{Introduction}

Quantum fields on noncommutative Moyal spacetime \cite{Douglas:2001ba,Szabo:2001kg} present an UV/IR mixing phenomenon \cite{Minwalla:1999px} which may prevent the field theory from being renormalizable. The obstruction to renormalizability is caused by certain interactions between virtual particles with high and low momenta which originate infinities that cannot be absorbed into a redefinition of the physical parameters. Still, in some noncommutative theories renormalizability can be recovered after an appropriate modification of the free-field propagator which takes into account Langmann-Szabo duality---an interchange between large and small energy scales~\cite{Langmann:2002cc}. In fact, some renormalization properties of these modified noncommutative theories get improved with respect to their commutative counterparts. This mechanism, discovered by H.\ Grosse and R.\ Wulkenhaar, has been applied to $\lambda\phi^4$ theory in four-dimensional Moyal spacetime~\cite{Grosse:2004yu}, where the perturbative renormalizability and the absence of a Landau pole were proved~\cite{Disertori:2006nq}.

After the success of this procedure for a self-interacting scalar field, the formulation of a renormalizable noncommutative gauge theory along this line has been studied \cite{deGoursac:2007gq,Grosse:2007dm,Blaschke:2007vc}. However, the joint implementation of the appropriately modified free propagator and gauge invariance in a full renormalizable theory has not been accomplished yet. This open problem currently draws attention to the study of noncommutative gauge theories.

In the present article we study $U_\star(1)$ gauge field theory---i.e. the generalization to Moyal spacetime of $U(1)$ gauge theory---from a worldline perspective, representing the trace of the gauge-fixed one-loop differential operator in terms of a particle path integral. The worldline formalism is a very efficient method to compute scattering amplitudes and other physical quantities in Quantum Field Theory~\cite{Schubert:2001he}. Recently, the use of worldline techniques in phase space has proved particularly convenient for dealing with nonlocal operators, which are distinctive in noncommutative theories \cite{Bonezzi:2012vr,Vinas:2014exa}.

One of the advantages of the worldline formalism to study gauge theories is that it is based in the background field method, so that gauge invariance is explicitly preserved and a considerable simplification with respect to the usual diagrammatic technique is obtained. We apply worldline techniques to derive a master formula for the one-loop effective action, from which $n$-point functions can be obtained. As an illustration, we compute the so-called planar and non-planar contributions to the photon self-energy. Planar terms, which contain all UV divergences of the theory, provide the $\beta$-function of $U_\star(1)$. Non-planar contributions, responsible for the UV/IR mixing, are shown to be given by terms containing nonlocal operators with both left- and right-Moyal multiplication or, equivalently, by Seeley-de Witt coefficients which cannot be expressed as the Moyal product of the fields~\cite{Vassilevich:2005vk}. We expect that the results presented in this article provide a useful tool for the perturbative study of noncommutative gauge theories, in particular, in the context of Grosse-Wulkenhaar models.

The article is organized as follows. In the remainder of this section we present the noncommutative Moyal product together with a few useful properties, and establish  our notation. In section \ref{u1} we shortly give the fundamentals of $U_\star(1)$ theory, whereas in section \ref{bfm} we apply the background field method to compute the relevant (nonlocal) operator of quantum fluctuations whose heat-trace determines the effective action of the theory. In section \ref{ht} we implement the worldline formalism in phase space to obtain a master formula for the effective action, which is presented in section \ref{mf}. In section \ref{mean} we explicitly compute the mean values using worldline techniques and make an analysis of the resulting Bern-Kosower form factors. After showing in section~\ref{1p} the vanishing of tadpole contributions, we study in section \ref{2p} the photon self-energy: we compute the $\beta$-funcion (section \ref{bf}), and the finite part of planar contributions (section \ref{pl}) as well as non-planar contributions (section \ref{npl}). Finally, in section \ref{conclu} we draw our conclusions. Appendices contain some material related to other types of worldline boundary conditions (appendix \ref{pbc}), divergences of 3- and 4-point functions (appendix \ref{3y4}) and some mathematical identities which are useful to prove that the photon polarization is transversal (appendix \ref{bessels}).

\subsection{The Moyal product}
Given two fields $\phi(x)$ and $\psi(x)$, with $x\in\mathbb{R}^4$ (Euclidean four-dimensional spacetime), we define the associative but noncommutative Moyal product
\begin{align}\label{mp}
  (\phi\star\psi)(x)=e^{i\,\partial_y \theta \partial_z}\ \phi(x+y)\,\psi(x+z)\,\big|_{y=z=0}\,,
\end{align}
where $\partial$ denotes the four-component gradient\footnote{To avoid cluttering we will most frequently omit the indices of matrices and four vectors; for instance, expression $\theta\partial$ represents $\theta_{\mu\nu}\partial_\nu$, $x\theta\partial$ represents $\theta_{\mu\nu}x_\mu\partial_\nu$, etc.} and $\theta$ a real antisymmetric matrix in $\mathbb{R}^{4\times 4}$ with dimensions of length squared, which we assume to be nondegenerate. The elements $\theta_{\mu\nu}$ of the noncommutativity matrix $\theta$ set a deformation of the usual commutative spacetime: under this $\star$-product the coordinates now satisfy the commutation relation $[x_\mu,x_\nu]:=x_\mu\star x_\nu-x_\nu\star x_\mu=2i\theta_{\mu\nu}$.

From definition \eqref{mp} one can formally derive the useful expressions
\begin{align}
  \phi\star\psi&=L(\phi)\,\psi=\phi(x+i\theta\partial)\,\psi\,,\\
          &=R(\psi)\,\phi=\psi(x-i\theta\partial)\,\phi\,,
\end{align}
where $L$ and $R$ denote left- and right-Moyal multiplication, respectively. It is sometimes convenient to make use of the representation of Moyal multiplication in Fourier space,
\begin{align}\label{mpinf}
  \mathcal{F}(\phi_1\star \phi_2\star \phi_3\star \ldots)(p)&=\int d\bar{p}_1d\bar{p}_2d\bar{p}_3\ldots
  \bar\delta\left(p_1+p_2+p_3+\ldots-p\right)\,\times\nonumber\\
  &\mbox{}\times\tilde \phi_1(p_1)\,\tilde \phi_2(p_2)\,\tilde \phi_3(p_3)\ldots\ e^{-i\sum_{i<j}p_i\theta p_j}\,.
\end{align}
Both the symbols $\mathcal{F}(\phi)$ and $\tilde{\phi}$ are used for the Fourier transform of a function $\phi$. The overlined $d\bar{p}$ means $d^4p/(2\pi)^4$; we also use an overline in Dirac delta functions to represent a factor $(2\pi)^4$, so that $\bar\delta=(2\pi)^4\,\delta$. Note that in Fourier space the effect of noncommutativity amounts to a phase---known as twisting factor---involving all products of momenta $p_i\theta p_j$. Since the twisting factor is not invariant under any permutation of momenta but only under cyclic ones, then each interaction vertex involved in a particular process in the commutative theory gives rise to many different inequivalent processes in the noncommutative theory. In Feynman diagram language, this means that a given diagram in commutative spacetime corresponds in Moyal spacetime to different contributions according to the non-cyclic interchanges of the fields attached to each vertex of the diagram. In accordance with all these possible interchanges, diagrams can be planar or non-planar and thus present very different physical consequences.

Let us finally mention that under the integral sign the following two properties hold:
\begin{align}
  \int_{\mathbb{R}^4}dx\ \phi\star \psi&=\int_{\mathbb{R}^4}dx\ \phi\,\psi\,,\\[2mm]
  \int_{\mathbb{R}^4}dx\ \phi\star \psi\star \chi&=\int_{\mathbb{R}^4}dx\ \chi\star\phi\star\psi\,.
\end{align}
The first property is a consequence of the fact that the difference between the Moyal product and the ordinary commutative product is a total derivative, and it implies that in noncommutative theories terms in the action that are quadratic in the fields do not need to involve Moyal product. The second property is an immediate consequence of the first one.

\section{$U_\star(1)$ gauge field theory}\label{u1}

There exists a mathematically rigorous formulation of classical noncommutative gauge theories \cite{Connes:1994yd,Landi:1997sh}. In this section we just introduce some basic concepts of pure $U_\star(1)$, the generalization of $U(1)$ gauge theory to noncommutative Moyal spacetime. As we will see, even in the absence of matter, the noncommutativity of Moyal spacetime introduces self-interactions for the photons; the resulting theory is, in many aspects, much like pure non-abelian Yang-Mills theory.

To begin, let us consider the function
\begin{align}
  U(x)=e_\star^{i\alpha(x)}=1+i\alpha(x)-\tfrac12\,\alpha(x)\star \alpha(x)+\ldots
\end{align}
whose Moyal inverse, $U\star U^{-1}=U^{-1}\star U=1$, is given by $U^{-1}=e_\star^{-i\alpha(x)}$. Such function defines a transformation of a gauge field $A_\mu(x)$ as
\begin{align}
  A_\mu(x)\rightarrow U\star A_\mu\star U^{-1}+i\,U\star \partial_\mu U^{-1}\,.
\end{align}
Consequently, the covariant derivative
\begin{align}
  D_\mu=\partial_\mu-iA_\mu
\end{align}
and the field strength
\begin{align}
  F_{\mu\nu}=i\,[D_\mu,D_\nu]=\partial_\mu A_\nu-\partial_\nu A_\mu
  -i\,[A_\mu,A_\nu]
\end{align}
(where, as before, $[A_\mu,A_\nu]=A_\mu\star A_\nu-A_\nu\star A_\mu$) transform covariantly under $U_\star(1)$, i.e., $D_\mu\rightarrow U\star D_\mu\star U^{-1}$ and $F_{\mu\nu}\rightarrow U\star F_{\mu\nu}\star U^{-1}$. With these ingredients we can now construct the following invariant action
\begin{align}\label{theaction}
  S[A]=\frac{1}{4e^2}\int_{\mathbb{R}^4}dx\ F_{\mu\nu}\star F_{\mu\nu}\,,
\end{align}
where $e^2$ is the bare coupling constant. Since $F_{\mu\nu}$ contains both linear and quadratic terms in the gauge field $A_\mu$, the action $S[A]$ involves cubic and quartic self-interactions for the photons. There is thus an evident similarity between noncommutative $U_\star(1)$ and commutative non-abelian Yang-Mills theories that, as we will see, manifests also in the quantization of the theory.

In Fourier space, the action reads
\begin{align}\label{theactioninF}
  S[A]&=\frac{1}{2e^2}\int d\bar\sigma
  \,A_\mu(\sigma)\,A_\nu(-\sigma)
  \left\{\delta_{\mu\nu}\,\sigma^2-\sigma_\mu\sigma_\nu\right\}+\mbox{}\nonumber\\[2mm]
  &-\frac{1}{e^2}\int d\bar\sigma_1d\bar\sigma_2d\bar\sigma_3\ \bar\delta\left(\sigma_1+\sigma_2+\sigma_3\right)
  \,A_\mu(\sigma_1)A_\nu(\sigma_2)A_\nu(\sigma_3)\times\mbox{}\nonumber\\[2mm]
  &\times \sigma_{3\mu}\left\{e^{i\sum_{i<j}\sigma_i\theta\sigma_j}-e^{-i\sum_{i<j}\sigma_i\theta\sigma_j}\right\}
  +\mbox{}\nonumber\\[2mm]
  &+\frac{1}{2e^2}\int d\bar\sigma_1d\bar\sigma_2d\bar\sigma_3d\bar\sigma_4
  \ \bar\delta\left(\sigma_1+\sigma_2+\sigma_3+\sigma_4\right)
  \,e^{i\sum_{i<j}\sigma_i\theta\sigma_j}\times\mbox{}\nonumber\\[2mm]
  &\times \left\{A_\mu(\sigma_1)A_\mu(\sigma_2)A_\nu(\sigma_3)A_\nu(\sigma_4)-
  A_\mu(\sigma_1)A_\nu(\sigma_2)A_\mu(\sigma_3)A_\nu(\sigma_4)\right\}
  \,.
\end{align}
The quadratic term describes a massless field with a transverse propagator. The cubic self-interaction corresponds to the term $\partial_\mu A_\nu\star[A_\mu,A_\nu]$ whereas the quartic one to the difference between the terms $A_\mu\star A_\mu \star A_\nu\star A_\nu$ and $A_\mu\star A_\nu \star A_\mu\star A_\nu$. Of course, for $\theta=0$ all interactions vanish and we are left with the usual free QED.

\section{The background field method}\label{bfm}

In order to study the one-loop effective action of $U_\star(1)$ we consider a fixed arbitrary background $A_\mu(x)$ and write the gauge field as $A_\mu(x)+a_\mu(x)$, so now the quantum fluctuations of the field are represented by $a_\mu(x)$. If we perform this shift in the action \eqref{theaction}, the terms which are quadratic in the quantum field read
\begin{align}\label{thequadraticaction}
  S^{(2)}=\frac{1}{2e^2}\int_{\mathbb{R}^4}dx\ \left\{
  -a_\mu[D_\nu,[D_\nu,a_\mu]]-[D_\mu,a_\mu][D_\nu,a_\nu]+2i\,a_\mu[F_{\mu\nu},a_\nu]
  \right\}\,,
\end{align}
where now the covariant derivative $D_\mu$ and the field strength $F_{\mu\nu}$ depend exclusively on the background field $A_\mu$. If, in addition, we choose the gauge condition $[D_\mu,a_\mu]=0$ and introduce the corresponding gauge fixing term (proportional to $[D_\mu,a_\mu]^2$) in the Feynman-'t Hooft gauge, then the second term in \eqref{thequadraticaction} cancels and the quadratic part of the action takes the simpler form
\begin{align}\label{thequadraticactionfixed}
  S^{(2)}_{\rm gauge}=\frac{1}{2e^2}\int_{\mathbb{R}^4}dx\ a_\mu\,\delta^2 S_{\rm gauge}\,a_\nu\,,
\end{align}
where the nonlocal operator $\delta^2 S_{\rm gauge}$ is given by
\begin{align}\label{dS2}
  \delta^2S_{\rm gauge}=-\delta_{\mu\nu}[D_\rho,[D_\rho,\cdot\,]]+2i\,[F_{\mu\nu},\cdot\,]\,.
\end{align}
This gauge choice is essential to get a minimal operator. In terms of left- and right-Moyal multiplications this operator can also be written as
\begin{align}\label{thegaugeoperator}
  \delta^2S_{\rm gauge}&=
  -\delta_{\mu\nu}\left\{\partial-iL(A)+iR(A)\right\}^2+2i\left\{L(F_{\mu\nu})-R(F_{\mu\nu})\right\}\nonumber\\
  &=-\delta_{\mu\nu}\left\{\partial-iA(x+i\theta\partial)+iA(x-i\theta\partial)\right\}^2+\mbox{}\nonumber\\
  &\mbox{}+2i\left\{F_{\mu\nu}(x+i\theta\partial)-F_{\mu\nu}(x-i\theta\partial)\right\}\,,
\end{align}
acting on four-component functions $a_\mu(x)\in \mathbb{R}^4\times L_2(\mathbb{R}^4)$. As already mentioned, the first term in \eqref{thegaugeoperator} is diagonal in the $\mu,\nu$-indices, but the second term has an internal structure that mixes these indices by means of the antisymmetric expression $F_{\mu\nu}$. The term ``gauge'' in $\delta^2S_{\rm gauge}$ is used to remark that it represents the quantum fluctuations of the gauge field and that it does not take into account the contributions of the ghost fields. However, for the chosen gauge, the corresponding ghost operator is simply given by
\begin{align}\label{theghostoperator}
  \delta^2S_{\rm ghost}&=-\left\{\partial-iL(A)+iR(A)\right\}^2
  \nonumber\\ &
  =-\left\{\partial-iA(x+i\theta\partial)+iA(x-i\theta\partial)\right\}^2\,,
\end{align}
acting on one-component Grassmann fields $\bar c(x),c(x)$. The ghost operator thus coincides with the diagonal part of the gauge operator (cfr.\ eq.~\eqref{thegaugeoperator}): this turns out to be quite helpful in the computation of the effective action below.

The one-loop effective action can then be expressed in terms of the functional determinant of these operators
\begin{align}
  \Gamma[A]=-\log{\rm Det}^{-\frac12}\left\{\delta^2S_{\rm gauge}\right\}
  -\log{\rm Det}\left\{\delta^2S_{\rm ghost}\right\}\,.
\end{align}
We finally regularize these determinants by means of the heat-traces of the quantum fluctuation operators
\begin{align}\label{theeffaction}
  \Gamma[A]=-\frac12\int_{\Lambda^{-2}}^\infty \frac{d\beta}{\beta}\,e^{-m^2\beta}
  \left({\rm Tr}\,e^{-\beta\,\delta^2S_{\rm gauge}}-2\,{\rm Tr}\,e^{-\beta\,\delta^2S_{\rm ghost}}\right)\,.
\end{align}
Note that we have introduced both an IR and an UV regulator $m$ and $\Lambda$ which prevent the integral to diverge at respectively large and small values of the Schwinger proper time $\beta$. Expression \eqref{theeffaction} shows one of the advantages of this formulation, namely that one can easily take account of the ghost contributions. Indeed, by means of the heat-trace, gauge contributions to the effective action arise from the exponentiation of the two terms in \eqref{thegaugeoperator}. If we expand this exponential in powers of the fields, there are terms which involve $F_{\mu\nu}$ (from the second term in \eqref{thegaugeoperator}), that are not present in the ghost heat-trace, and terms which are constructed only from powers of the first term, which do appear also in the ghost part. In the gauge heat-trace, these latter terms are multiplied by 4---due to the trace in the $\mathbb{R}^4$ part of $\mathbb{R}^4\times L_2(\mathbb{R}^4)$---whereas in the ghost part the same terms are just multiplied by the $-2$ coefficient of expression \eqref{theeffaction}.  Hence, we can consider only the gauge part, multiplying by a factor 2 those terms which do not involve $F_{\mu\nu}$ and leaving unmodified those terms which contain $F_{\mu\nu}$; once this prescription is followed, the ghost contribution is automatically incorporated.

\section{Worldline determination of the heat-trace}\label{ht}

In this section we determine the heat-trace ${\rm Tr}\,e^{-\beta\,\delta^2S_{\rm gauge}}$ of the nonlocal operator $\delta^2S_{\rm gauge}$ using the worldline formalism. In terms of phase space path integrals the trace can be written as
\begin{align}
  {\rm Tr}\,e^{-\beta\,\delta^2S_{\rm gauge}}&=
  {\rm tr}\int_{\mathbb{R}^4}dx\ \langle x|e^{-\beta\,\delta^2S_{\rm gauge}}|x\rangle\nonumber\\
  &={\rm tr}\int_{\mathbb{R}^4}dx\ \int\mathcal{D}x(t)\mathcal{D}p(t)
  \ e^{-\int_0^\beta dt\,\left\{-ip(t)\dot{x}(t)+\delta^2S_W(x(t),p(t))\right\}}\,,
\end{align}
where the trajectories $x(t)$ satisfy $x(0)=x(\beta)=x$. The expression ``$\rm tr$'' denotes the trace over $\mu,\nu$-indices in the $\mathbb{R}^4$ part of $\mathbb{R}^4\times L_2(\mathbb{R}^4)$. The function $\delta^2S_W(x(t),p(t))$ is obtained by replacing $x\rightarrow x(t)$ and $\partial\rightarrow ip(t)$ in the Weyl-ordered expression of the operator $\delta^2S_{\rm gauge}$. Weyl ordering is required by the midpoint prescription in the time-slicing definition of the path integral. Nevertheless one can show from a formal Taylor expansion that for any pair of functions $\phi,\psi$ the operators $\phi(x+i\theta \partial)$, $\psi(x-i\theta \partial)$, the mixed product $\phi(x+i\theta \partial)\cdot \psi(x-i\theta \partial)$ and also the symmetrized expressions $\partial\cdot \phi(x+i\theta \partial)+\phi(x+i\theta \partial)\cdot\partial$ and $\partial\cdot \psi(x-i\theta \partial)+\psi(x-i\theta \partial)\cdot\partial$ are already Weyl-ordered; this means that no extra terms are needed in order to write them as completely symmetrized expressions of the the operators $x$ and $\partial$.
Hence,
\begin{align}\label{weyl}
  \delta^2S_W(x,p)&=\left\{p+A(x+\theta p)-A(x-\theta p)\right\}^2+\mbox{}\nonumber\\
  &\mbox{}-2i\left\{F_{\mu\nu}(x+\theta p)-F_{\mu\nu}(x-\theta p)\right\}\,.
\end{align}
Note that the first term contains squares of the gauge field which must be read as conventional Moyal squares $(A^2_\star)(y)$, evaluated at the operators $y=x\pm \theta p$, i.e.\ $(A^2_\star)(x\pm \theta p)$ are regular functions of operators $x\pm \theta p$. Therefore, by expressing them in terms of their Taylor expansions, one can promptly check that they are written in Weyl-ordered form---see discussion in~\cite{Bonezzi:2012vr} for further details.

It is now convenient to rescale the proper time parameter as $t\rightarrow \beta t$ and to redefine the trajectories as $x(t)\rightarrow x+\sqrt{\beta}\,x(t)$ and $p(t)\rightarrow p(t)/\sqrt{\beta}$ in terms of dimensionless functions of the rescaled proper time $t\in[0,1]$. Note that the projection of the trajectories onto the configuration space now represents perturbations around the fixed position $x$ so that $x(t)$ satisfies homogeneous Dirichlet conditions, $x(0)=x(1)=0$. After these redefinitions the heat-trace can be written as
\begin{align}\label{thetrace0}
  &{\rm Tr}\,e^{-\beta\,\delta^2S_{\rm gauge}}
  =\nonumber\\
  &=\mathcal{N}(\beta)\,{\rm tr}\int_{\mathbb{R}^4}dx\ \left\langle
  \ e^{-\beta\int_0^1 dt\,\left\{
  \frac{2}{\sqrt{\beta}}\,p(t)\,[A(+)-A(-)]+[A(+)-A(-)]^2-2i\,[F(+)-F(-)]
  \right\}}\right\rangle\,,
\end{align}
where the signs $(\pm)$ indicate that the field must be evaluated at $x+\sqrt{\beta}\,x(t)\pm \theta p(t)/\sqrt{\beta}$, and $F$ represents the tensor field $F_{\mu\nu}$. The mean value in~\eqref{thetrace0} is defined as
\begin{align}
  \left\langle\,\ldots\,\right\rangle=
  \frac{\displaystyle \int\mathcal{D}x(t)\mathcal{D}p(t)
  \ e^{-\int_0^1 dt\,\left\{p^2-ip\dot{x}\right\}}
  \ \ldots}
  {\displaystyle \int\mathcal{D}x(t)\mathcal{D}p(t)
  \ e^{-\int_0^1 dt\,\left\{p^2-ip\dot{x}\right\}}}\,,
\end{align}
where the normalization has been chosen so that $\langle 1\rangle=1$. The subsequent normalization factor $\mathcal{N}(\beta)$ can be determined from the value of the heat-trace in the free case. Indeed, for $A_\mu(x)=0$ we obtain
\begin{align}
  {\rm Tr}\,e^{-\beta\,\delta^2S_{\rm gauge}}
  &=\mathcal{N}(\beta)\,{\rm tr}\int_{\mathbb{R}^4}dx\ \left\langle 1 \right\rangle\nonumber\\[2mm]
  &={\rm Tr}\,e^{-\beta\,(-\partial)^2}
  =\int_{\mathbb{R}^4}dx\ \frac{1}{(4\pi\beta)^2}\ 4\,.
\end{align}

In conclusion, we obtain for the heat-trace
\begin{align}\label{thetrace}
  &{\rm Tr}\,e^{-\beta\,\delta^2S_{\rm gauge}}
  =\nonumber\\
  &=\frac{1}{(4\pi\beta)^2}\,{\rm tr}\int_{\mathbb{R}^4}dx\ \left\langle
  \ e^{-\beta\int_0^1 dt\,\left\{
  \frac{2}{\sqrt{\beta}}\,p(t)\,[A(+)-A(-)]+[A(+)-A(-)]_\star^2-2i\,[F(+)-F(-)]
  \right\}}\right\rangle\,.
\end{align}
The $\star$-symbol reminds us that, as discussed below eq.\ \eqref{weyl}, the squares correspond to Moyal multiplication.

\section{The effective action}\label{mf}

We have now all the ingredients to write down an expression for the effective action from which $n$-point functions can be computed in terms of worldline mean values. Replacing \eqref{thetrace} into \eqref{theeffaction} we obtain, after expanding the exponential,
\begin{align}\label{masteraction}
  &\Gamma[A]=-\frac12 \int_{\Lambda^{-2}}^\infty
  \frac{d\beta}{\beta}\,\frac{e^{-m^2\beta}}{(4\pi \beta)^2}
  \ \sum_{n=1}^\infty(-\beta)^{n}
  \int_0^1 dt_1\int_0^{t_1}dt_2\ldots\int_0^{t_{n-1}}dt_n
  \,\times\mbox{}\nonumber\\[2mm]
  &\mbox{}\times
  \tilde{\rm tr}\int_{\mathbb{R}^4} dx\left\langle
  \ \prod_{i=1}^n\left(
  \frac{2}{\sqrt{\beta}}\,p_\mu(t_i)\,V_\mu^A(t_i)+V^{AA}(t_i)-2i\,V_{\mu\nu}^F(t_i)\right)
  \right\rangle\,,
\end{align}
where we have defined the vertex functions
\begin{align}
  V_\mu^A(t)&=A_\mu(+)-A_\mu(-)\,,\\[2mm]
  V^{AA}(t)&=(A_\mu\star A_\mu)(+)+(A_\mu\star A_\mu)(-)-2\,A_\mu(+)\,A_\mu(-)\,,\\[2mm]
  V_{\mu\nu}^F(t)&=F_{\mu\nu}(+)-F_{\mu\nu}(-)\,.
\end{align}
The signs $\pm$ between brackets indicate that the field is evaluated at $x+\sqrt{\beta}\,x(t)\pm\theta p(t)/\sqrt{\beta}$, respectively. The symbol $\tilde{\rm tr}$ means that for the diagonal terms in the expansion (i.e.\ for those that do not contain $V_{\mu\nu}^F$) the trace amounts to a simple multiplicative factor of 2: this automatically takes into account of the ghost part of the heat-trace, as explained at the end of section \ref{bfm}. Note also that the term $n=0$, being field-independent, has been omitted.

Expression \eqref{masteraction} allows the successive computation of $n$-point functions according to the number of powers of the gauge field, taking into account that the vertex $V_\mu^A$ is linear in $A_\mu$, $V^{AA}$ is quadratic, and $V_{\mu\nu}^F$ contains both linear and quadratic terms in $A_\mu$. Whenever a term in a product of vertices contains fields evaluated either at $(+)$-type or $(-)$-type arguments only, its contribution to the effective action corresponds to a planar Feynman diagram. On the contrary, products of fields evaluated at different types of arguments give the contributions of non-planar diagrams.

In the next section we will use the worldline approach to indicate how to compute the phase space mean value in this master formula.

\section{Mean values}\label{mean}

The master formula \eqref{masteraction} requires the calculation of mean values of the form
\begin{align}\label{interm}
  \int_{\mathbb{R}^4} dx\,
  \left\langle
  \ \prod_{i=1}^n V_i\left(x+\sqrt{\beta}\,x(t_i)\pm_i \theta p(t_i)/\sqrt{\beta}\right)
  \right\rangle\,,
\end{align}
where each of the functions $V_1,V_2,\ldots,V_n$ represents a field contained either in $V^{AA}$ or $V_{\mu\nu}^F$; later in this section we will consider mean values which also contain fields from the vertex $V_\mu^A$. Recall that the double sign $\pm_i$ indicates that the field $V_i$ comes from the right- or left-Moyal multiplication, respectively. In terms of Fourier transforms this quantity can be written as
\begin{align}\label{interm2}
  &\int_{\mathbb{R}^{4\times n}}d\bar\sigma_1\ldots d\bar\sigma_n
  \ \bar\delta({\textstyle \sum}\sigma_i)
  \ \tilde V_1(\sigma_1)\ldots \tilde V_n(\sigma_n)
  \times\mbox{}\nonumber\\[2mm]
  &\mbox{}\times
  \left.\left\langle
  e^{i\sum_{i=1}^n\left(\sqrt{\beta}\,x(t_i)\sigma_i
  +\frac{1}{\sqrt{\beta}} p(t_i)\rho_i\right)}\right
  \rangle\right|_{\rho_i=\mp_i\theta\sigma_i}\,,
\end{align}
where the delta function that enforces the conservation of the total momentum is due to the integral over the ``zero-mode'' $x$. The mean value in the presence of arbitrary external sources $j(t),k(t)$ coupled to the fields $x(t),p(t)$ can be readily computed after the standard procedure of completing squares and inverting the differential operator of the resulting quadratic form. The result reads
\begin{align}\label{meanvalue}
  &\left\langle\,e^{i\int_0^1 dt\left\{k(t)p(t)+j(t)x(t)\right\}}\,\right\rangle
  =\frac{ \int\mathcal{D}x(t)\mathcal{D}p(t)\ e^{-\int_0^1 dt\,\left\{p^2-ip\dot{x}\right\}}
  \,e^{i\int_0^1 dt\left\{kp+jx\right\}}}
  {\int\mathcal{D}x(t)\mathcal{D}p(t)\ e^{-\int_0^1 dt\,\left\{p^2-ip\dot{x}\right\}}}\nonumber\\
  &=\exp{\left(-\int\int dtdt'\left\{\tfrac14\,k(t)k(t')+g(t,t')j(t)j(t')
  +\tfrac{i}{2}\, h(t,t')k(t)j(t')\right\}\right)}
  \,,
\end{align}
where $h(t,t')$ and $g(t,t')$ are elements of the Green's matrix
\begin{align*}
D^{-1}(t,t') =\left(
\begin{array}{cc}
\frac12 & \frac{i}{2} h(t,t')\\[2mm]
\frac{i}{2}h(t',t) & 2g(t,t')
\end{array} \right)\,,
\end{align*}
which is the inverse of the phase space kinetic operator with $x(t)$ satisfying homogeneous Dirichlet boundary conditions, i.e.
\begin{align*}
g(t,t') & =-\frac12|t-t'|-tt'+\frac12(t+t')\,,\quad \\
h(t,t') &= 2\partial_t g(t,t') = -\epsilon(t-t')-2t'+1~.
\end{align*}
However, as shown in the appendix~\ref{pbc}, expression~\eqref{interm2}, and ultimately the effective action, can be equivalently computed using the homogeneous trans\-la\-tionally-invariant Green's function
\begin{align}
  G(t-t')&:=-\frac12|t-t'|+\frac12(t-t')^2\,,\\
  \quad H(t-t')&:= 2\dot G(t-t')=-\epsilon(t-t')+2(t-t')\,,
\end{align}
where $\epsilon(\cdot)$ is the sign function.
In turn this thus leads to the phase space worldline propagators
\begin{align*}
& \langle p_\mu(t) p_\nu(t')\rangle =\delta_{\mu\nu}~\frac12\,, \\
& \langle x^\mu(t) x^\nu(t')\rangle =\delta^{\mu\nu}~2G(t-t')\,, \\
& \langle p_\mu(t) x^\nu(t') \rangle =\delta_\mu^\nu~ i\dot  G(t-t') ~.
\end{align*}
These propagators are the phase space counterparts of the ``string-inspired'' configuration space  propagator adopted in~\cite{Schubert:2001he}, and correspond to an alternative way of factoring out the zero mode of the kinetic operator (see also~\cite{Bastianelli:2003bg} for a discussion on the factorization of the worldline zero mode).  Basically, in the expression~\eqref{interm2}, terms that involve the difference between the Dirichlet Green's function and the string-inspired one are proportional to the total four-momentum, and hence vanish.

The mean value in expression~\eqref{interm2} can now be computed by replacing
\begin{align}
  k(t)&:=\beta^{-\frac12} \sum_{i=1}^n \delta(t-t_i)\,\rho_i\,,\\
  j(t)&:=\beta^{\frac12} \sum_{i=1}^n \delta(t-t_i)\,\sigma_i
\end{align}
into expression \eqref{meanvalue}; the result reads
\begin{align}\label{themeanvalue}
  \left\langle
  e^{i\sum_{i=1}^n\left(\sqrt{\beta}\,x(t_i)\sigma_i
  +\frac{1}{\sqrt{\beta}} p(t_i)\rho_i\right)}\right\rangle
  =e^{-\sum_{i,j} \left\{\frac1{4\beta}\,\rho_i\rho_j
  +\beta\,G_{ij}\,\sigma_i\sigma_j+i \dot G_{ij}\,\rho_i\sigma_j\right\}}
  \,,
\end{align}
where $G_{ij}:=G(t_i-t_j)$ and $\dot G_{ij}:=\dot G(t_i-t_j)$. Finally, expression \eqref{interm} can now be written as
\begin{align}\label{expval}
  &\int_{\mathbb{R}^4} dx
  \left\langle
  \prod_{i=1}^n V_i\left(x+\sqrt{\beta}\,x(t_i)\pm_i \theta p(t_i)/\sqrt{\beta}\right)
  \right\rangle=\nonumber\\
  &=\int_{\mathbb{R}^{4\times n}}d\bar\sigma_1\ldots d\bar\sigma_n
  \,\bar\delta(\sigma_1+\ldots+\sigma_n)
  \,\tilde V_1(\sigma_1)\ldots \tilde V_n(\sigma_n)
  \times\mbox{}\nonumber\\[2mm]
  &\mbox{}\times
  e^{-\sum_{i,j} \left\{\frac1{4\beta}\,(\mp_i\theta\sigma_i)(\mp_j\theta\sigma_j)
  +\beta\,G_{ij}\,\sigma_i\sigma_j+i\dot G_{ij}\,(\mp_i\theta\sigma_i)\sigma_j\right\}}\,.
\end{align}

Expression \eqref{masteraction} also requires the calculation of mean values including the functions $p_\mu(t_i)\,V_\mu^A(t_i)$ for some $t_i$,
\begin{align}\label{intermpe}
  \int_{\mathbb{R}^4} dx
  \left\langle \frac{2}{\sqrt{\beta}}\,p(t_a)\ \frac{2}{\sqrt{\beta}}\,p(t_b)\ldots
  \ \prod_{i=1}^n V_i\left(x+\sqrt{\beta}\,x(t_i)\pm_i \theta p(t_i)/\sqrt{\beta}\right)
  \right\rangle\,,
\end{align}
where $t_a,t_b,\ldots$ belong to the set $\{t_1,\ldots,t_n\}$; the corresponding fields are then $V_a=A_\mu(x+\sqrt{\beta}\,x(t_a)\pm_a \theta p(t_a))$ and must be Lorentz-contracted with $p_\mu(t_a)$. In terms of Fourier transforms this quantity can be written as
\begin{align}\label{interm2pe}
  &\int_{\mathbb{R}^{4\times n}}d\bar\sigma_1\ldots d\bar\sigma_n
  \,\bar\delta(\sigma_1+\ldots+\sigma_n)
  \,\tilde V_1(\sigma_1)\ldots \tilde V_n(\sigma_n)
  \times\mbox{}\nonumber\\[2mm]
  &\mbox{}\times
  (-2i)\,\frac{\partial}{\partial\rho_a}
  \ (-2i)\,\frac{\partial}{\partial\rho_b}\ldots
  \left.\left\langle
  e^{i\sum_{i=1}^n\left(\sqrt{\beta}\,x(t_i)\sigma_i
  +\frac{1}{\sqrt{\beta}} p(t_i)\rho_i\right)}\right
  \rangle\right|_{\rho_i=\mp_i\theta\sigma_i}\,,
\end{align}
so that expression \eqref{intermpe} results
\begin{align}\label{expvalp}
  &\int_{\mathbb{R}^4} dx
  \left\langle \frac{2}{\sqrt{\beta}}\,p(t_a)\ \frac{2}{\sqrt{\beta}}\,p(t_b)\ldots
  \ \prod_{i=1}^n V_i\left(x+\sqrt{\beta}\,x(t_i)\pm_i \theta p(t_i)/\sqrt{\beta}\right)
  \right\rangle
  =\nonumber\\
  &=\int_{\mathbb{R}^{4\times n}}d\bar\sigma_1\ldots d\bar\sigma_n
  \,\bar\delta(\sigma_1+\ldots+\sigma_n)
  \,\tilde V_1(\sigma_1)\ldots \tilde V_n(\sigma_n)
  \times\mbox{}\nonumber\\[2mm]
  &\mbox{}\times
  (-2i)\,\frac{\partial}{\partial\rho_a}
  \ (-2i)\,\frac{\partial}{\partial\rho_b}\ldots
  \left.e^{-\sum_{i,j} \left\{\frac1{4\beta}\,\rho_i\rho_j
  +\beta\,G_{ij}\,\sigma_i\sigma_j+i\dot G_{ij}\,\rho_i\sigma_j\right\}}
  \right|_{\rho_i=\mp_i\theta\sigma_i}\,.
\end{align}
Note that, though not explicitly indicated, the fields $\tilde{V}_a(\sigma_a),\tilde{V}_b(\sigma_b),\ldots$ contain Lorentz indices that must be contracted with the Lorentz indices in the gradients $\partial_{\rho_a},\partial_{\rho_b},\ldots$, respectively.

Before concluding this section, let us make some comments regarding the Bern-Kosower form factor
\begin{align}\label{bkf}
  e^{-\sum_{i,j} \left\{\frac1{4\beta}\,(\mp_i\theta\sigma_i)(\mp_j\theta\sigma_j)
  +\beta\,G_{ij}\,\sigma_i\sigma_j+i\dot G_{ij}\,(\mp_i\theta\sigma_i)\sigma_j\right\}}\,,
\end{align}
that appears both in expression~\eqref{expval} and \eqref{expvalp}. The term
\begin{align}\label{ff1}
  e^{-\beta\sum_{i,j} G_{ij}\,\sigma_i\sigma_j}
\end{align}
is the resulting form factor for the commutative case, $\theta=0$. After the usual small-$\beta$ expansion--- equivalent to a large-mass expansion---one obtains successive integer powers of $\sigma_i\sigma_j$, which represent higher-order derivatives of the fields in the effective action.

Let us next consider the term\footnote{Recall that, due to time-ordering in the Feynman path integral, $t_i>t_j$ for $i<j$.}
\begin{align}\label{ff2}
  e^{-i\,\sum_{i,j} \dot G_{ij}\,(\mp_i\theta\sigma_i)\sigma_j}
  =e^{i\,\sum_{i<j} \left[(\pm_i)+(\pm_j)\right]
  \,\left[\frac12-(t_i-t_j)\right]\,\sigma_i\theta\sigma_j}\,,
\end{align}
which is independent of $\beta$. Note that terms in the exponent involving indices $i,j$ for which $\pm_i\neq\pm_j$ (i.e., corresponding to vertices $V_i,V_j$ that act one by left- the other by right-multiplication) vanish. In other words, the sum in the exponent of expression \eqref{ff2} only involves pair of momenta of vertices which act both by left- or both by right-Moyal multiplication. Moreover, if all indices $i$ have the same sign $\pm_i$ (i.e., for a planar contribution), due to momentum conservation, expression~\eqref{ff2} reads
\begin{align}\label{fftf}
  e^{\pm_i\,i\,\sum_{i<j} \sigma_i\theta\sigma_j}\,,
\end{align}
which is the so-called twisting factor that gives the Moyal product of the vertices (see eq.\ \eqref{mpinf}). This means that for planar contributions, the full consequence of the term linear in $\theta$ in the exponent of expression \eqref{bkf} is to provide the $\star$-product of the fields. If $\pm_i=-$, i.e.\ if all fields act by left-multiplication, then the $\star$-product must be written according to time-ordering; if $\pm_i=+$, then the $\star$-product is reversed.

Finally, momentum conservation implies that the term
\begin{align}\label{ff3}
  e^{-\frac1{4\beta}\,\sum_{i,j} (\mp_i\theta\sigma_i)(\mp_j\theta\sigma_j)}
  =e^{-\frac1{4\beta}\left(\sum_{i=1}^n(\mp_i)\theta\sigma_i\right)^2}
\end{align}
gives no contribution if all indices $i$ have the same sign $\pm_i$. Otherwise, assume that some momenta---say $\sigma_1,\ldots,\sigma_l$ with $0<l<n$---correspond to vertices acting with left-Moyal product; then expression~\eqref{ff3} reads
\begin{align}\label{npreg}
  e^{-\frac1{\beta}\left|\theta\sum_{i=1}^l\sigma_i\right|^2}\,.
\end{align}
This is then a purely non-planar contribution, which appears as long as both left- and right-Moyal products are present.

In conclusion, for planar contributions one only gets the twisting factors \eqref{fftf}, which introduce time-ordered or reversed time-ordered Moyal products of the fields, and the phase \eqref{ff1}, which gives the usual series of higher-derivatives of the fields that is also present in the commutative case. On the contrary, non-planar contributions also contain terms as \eqref{npreg} which decrease faster than any power of $\beta$ as $\beta\rightarrow \infty$ and thus provide an UV-regularization of the effective action. However, as seen from expression \eqref{npreg}, this regularization has no effect if the sum of momenta $\sigma_1+\ldots+\sigma_l$ vanishes, so the original UV divergence turns into an IR divergence (UV/IR mixing).

An alternative approach to compute~\eqref{interm} is to Taylor expand the vertex functions $V_i$ around the zero-mode $x$, i.e. $V_i(x+z(t_i)) =e^{z(t_i)\cdot \partial_i} V_i(x)$, so that one gets
\begin{align}
& \int_{\mathbb{R}^4} dx \left\langle e^{\sum_i(\sqrt\beta x(t_i)\mp_i\frac{1}{\sqrt\beta} p(t_i))\cdot \partial_i}\right\rangle V_1(x_1)\cdots V_n(x_n)\Big\vert_{x_1=\cdots=x_n=x}\nonumber\\
&= \int_{\mathbb{R}^4} dx~ e^{\sum_{i,j}\left\lbrace \frac{1}{4\beta}(\mp_i)(\mp_j)+\beta G_{ij} +(\mp_i)\dot G_{ij} \right\rbrace\partial_i\cdot\partial_j}V_1(x_1)\cdots V_n(x_n)\Big\vert_{x_1=\cdots=x_n=x}\,,
\end{align}
where $\partial_i$ is the gradient of the $i$-th vertex.
This method might appear advantageous if one wants to write the final results in configuration space as it avoids a passage to Fourier space. On the other hand in Fourier space it turns out to be relatively easier to spot vanishing terms, and nonlocal contributions to the effective action involve functions of the Fourier momenta rather than functions of the derivatives of fields. It thus appears more natural in our context to work in Fourier space.

\section{One-point function}\label{1p}

As a first example, let us compute the one-point function to show that there are no tadpole contributions. Expression~\eqref{masteraction} indicates that the part of the effective action which is linear in the gauge field is given by the first and the third terms in the mean value---i.e., linear in $V^A_\mu$ and $V^F_{\mu\nu}$, respectively---in the contribution corresponding to $n=1$. In fact, since $F_{\mu\nu}$ is traceless, we would only obtain a contribution from the former. However, applying expression \eqref{expvalp} for $n=1$, we get
\begin{align}\label{1point}
  \int_{\mathbb{R}^4} dx\left\langle
  \frac{2}{\sqrt{\beta}}\,p_\mu(t)\,A_\mu(\pm)\right\rangle
  &=\int_{\mathbb{R}^{4}}d\bar\sigma\,\bar\delta(\sigma)\,\tilde A_\mu(\sigma)
  \ (-2i)\,\frac{\partial}{\partial\rho_{\mu}}
  \left.e^{-\frac1{4\beta}\,\rho^2}\right|_{\rho=\mp\theta\sigma}\nonumber\\[2mm]
  &=0\,.
\end{align}
The whole one-point function thus vanishes, and then an expansion around the trivial vacuum would be justified. Nevertheless, we will see that this configuration becomes unstable when non-planar contributions to the self-energy are considered.

\section{Two-point function}\label{2p}

In this section we study the one-loop two-point function. Let us consider first the planar contribution to the quadratic effective action $\Gamma^{(2)}$ arising from the mean value $\langle V^F_{\mu\nu}(t_1)V^F_{\mu\nu}(t_2)\rangle$ in expression \eqref{masteraction} for $n=2$,
\begin{align}\label{gamaf}
  &\frac{1}{8\pi^2}\int_{\Lambda^{-2}}^\infty
  \frac{d\beta}{\beta}\,e^{-m^2\beta}
  \int_0^1 dt_1\int_0^{t_1}dt_2
  \int_{\mathbb{R}^4} dx\left\langle V_{\mu\nu}^F(t_1)\,V_{\nu\mu}^F(t_2)\right\rangle\,.
\end{align}
Of course, since this term is quadratic in $F_{\mu\nu}$ it contains terms which are quadratic in $A_\mu$ but also cubic and quartic terms in the gauge field. Using eq.\ \eqref{expval} we compute the planar part $\Gamma^{(2)}_F$  of this expression, which receives equal contributions from the purely left- and purely right-Moyal operators,
\begin{align}\label{div3rd}
  \Gamma^{(2)}_F&=-\frac{1}{4\pi^2}\int_{\Lambda^{-2}}^\infty
  \frac{d\beta}{\beta}\,e^{-m^2\beta}
  \int_0^1 dt_1\int_0^{t_1}dt_2
  \int d\bar\sigma_1 d\bar\sigma_2\,\bar\delta(\sigma_1+\sigma_2)
  \times\mbox{}\nonumber\\[2mm]
  &\mbox{}\times
  \tilde F_{\mu\nu}(\sigma_1)\tilde F_{\mu\nu}(\sigma_2)
  \ e^{-2\beta\,G_{12}\,\sigma_1\sigma_2}\nonumber\\
  &=-\frac1{4\pi^2}
  \int d\bar\sigma\,\tilde F_{\mu\nu}(\sigma)\tilde F_{\mu\nu}(-\sigma)
  \int_{\Lambda^{-2}}^\infty
  d\beta\,\frac{e^{-m^2\beta}}{\beta}\int_0^1 dt\,t\,e^{-\beta\,t(1-t)\,\sigma^2}\,.
\end{align}
We have used momentum conservation and the vanishing of $G_{ij}$ at coincident times. The integral over the Schwinger time $\beta$ yields
\begin{align}
\Gamma^{(2)}_F= -\frac{1}{4\pi^2} \int d\bar\sigma\, \tilde F_{\mu\nu}(\sigma)  \tilde F_{\mu\nu}(-\sigma) \int_0^1 dt\, t\, \Gamma\left(0,\frac{m^2 +t(1-t)\,\sigma^2 }{\Lambda^2} \right)\,.
\end{align}
The UV-divergent part of such contribution can be obtained by extracting the $O(\beta^0)$ term in the last exponential in~\eqref{div3rd}. Hence
\begin{align}\label{FFdiv}
  \Gamma^{(2)}_F
  =-\frac1{8\pi^2}\,\Gamma(0,m^2/\Lambda^2)\int_{\mathbb{R}^4} dx
  \ F_{\mu\nu}(x)\star F_{\mu\nu}(x)+\ldots\,,
\end{align}
where the dots represent UV-finite contributions.

On the other hand, the contribution to the quadratic effective action arising from the mean value $\langle V^{AA}(t)\rangle$ in expression \eqref{masteraction}, for $n=1$, reads
\begin{align}\label{gamaaa}
  \frac{1}{16\pi^2} \int_{\Lambda^{-2}}^\infty
  \frac{d\beta}{\beta^2}\,e^{-m^2\beta}
  \int_0^1 dt
  \int_{\mathbb{R}^4} dx\left\langle V^{AA}(t)\right\rangle
  \,.
\end{align}
Using again eq.\ \eqref{expval}, we obtain the planar part $\Gamma^{(2)}_{AA}$ of this contribution,
\begin{align}\label{div2nd}
  \Gamma^{(2)}_{AA}&=\frac{1}{8\pi^2} \int_{\Lambda^{-2}}^\infty
  \frac{d\beta}{\beta^2}\,e^{-m^2\beta}
  \int d\bar\sigma\,\bar\delta(\sigma)
  \,\tilde A^2_\star(\sigma)\nonumber\\
  &=\frac{m^2}{8\pi^2}\,\Gamma(-1,m^2/\Lambda^2)
  \int_{\mathbb{R}^4}dx\,A_\star^2(x)
  \,,
\end{align}
This contribution would introduce a mass term for the gauge field, which diverges as $\sim \Lambda^2$ in the UV limit. However, one more contribution remains to be computed which will cancel the above quadratic divergence. This last contribution to $\Gamma^{(2)}$ corresponds to the mean value $\langle p_\mu(t_1)p_\nu(t_2)V_\mu^{A}(t_1)V_\nu^{A}(t_2)\rangle$ in \eqref{masteraction} for $n=2$,
\begin{align}\label{gamaa}
  &-\frac{1}{16\pi^2} \int_{\Lambda^{-2}}^\infty
  \frac{d\beta}{\beta}\,e^{-m^2\beta}
  \int_0^1 dt_1\int_0^{t_1}dt_2
  \,\times\mbox{}\nonumber\\[2mm]
  &\mbox{}\times
  \int_{\mathbb{R}^4} dx\left\langle
  \frac{2}{\sqrt{\beta}}\,p_\mu(t_1)\,V_\mu^A(t_1)\,\frac{2}{\sqrt{\beta}}\,p_\mu(t_2)\,V_\mu^A(t_2)
  \right\rangle\,.
\end{align}
The planar part $\Gamma^{(2)}_A$ of this contribution, using now eq.\ \eqref{expvalp}, is given by
\begin{align}
  \Gamma^{(2)}_A&=\frac{1}{4\pi^2} \int_{\Lambda^{-2}}^\infty
  \frac{d\beta}{\beta}\,e^{-m^2\beta}
  \int_0^1 dt_1\int_0^{t_1}dt_2
  \,\times\mbox{}\nonumber\\[2mm]
  &\mbox{}\times
  \int d\bar\sigma_1 d\bar\sigma_2\,\bar\delta(\sigma_1+\sigma_2)
  \,\tilde A_\mu(\sigma_1) \tilde A_\nu(\sigma_2)
  \times\mbox{}\nonumber\\[2mm]
  &\mbox{}\times
  \frac{\partial}{\partial\rho_{1\mu}}
  \,\frac{\partial}{\partial\rho_{2\nu}}
  \left.e^{-\sum_{i,j} \left\{\frac1{4\beta}\,\rho_i\rho_j
  +\beta\,G_{ij}\,\sigma_i\sigma_j+i\dot G_{ij}\,\rho_i\sigma_j\right\}}
  \right|_{\rho_i=-\theta\sigma_i}+\left(\theta\rightarrow-\theta\right)\,.
\end{align}
The last term in this expression represents the contribution of the operators acting by right-multiplication. However, both contributions coincide and give
\begin{align}\label{div1st}
  \Gamma^{(2)}_A&=\frac{1}{2\pi^2} \int_{\Lambda^{-2}}^\infty
  \frac{d\beta}{\beta}\,e^{-m^2\beta}
  \int_0^1 dt_1\int_0^{t_1}dt_2
  \,\times\mbox{}\nonumber\\[2mm]
  &\mbox{}\times
  \int d\bar\sigma\,\tilde A_\mu(\sigma)\tilde A_\nu(-\sigma)
  \ e^{2\beta\,G_{12}\,\sigma^2}
  \left\{-\frac{1}{2\beta}\,\delta_{\mu\nu}-\dot G^2_{12}\,\sigma_\mu\sigma_\nu\right\}
  \,.
\end{align}
After an appropriate expansion of the exponential $e^{2\beta\,G_{12}\,\sigma^2}$ for small $\beta$ we compute the divergent part of this expression,
\begin{align}
  \Gamma^{(2)}_A&=\frac{1}{2\pi^2} \int_{\Lambda^{-2}}^\infty
  \frac{d\beta}{\beta}\,e^{-m^2\beta}
  \int_0^1 dt_1\int_0^{t_1}dt_2
  \int d\bar\sigma\,\tilde A_\mu(\sigma)\tilde A_\nu(-\sigma)
  \,\times\mbox{}\nonumber\\[2mm]
  &\mbox{}\times
  \left\{-\frac{1}{2\beta}\,\delta_{\mu\nu}\left(1+2\beta\,G_{12}\,\sigma^2+\ldots\right)
  -\dot G^2_{12}\,\sigma_\mu\sigma_\nu\left(1+\ldots\right)\right\}\nonumber\\
  &=-\frac{m^2}{8\pi^2}\,\Gamma(-1,m^2/\Lambda^2)
  \int_{\mathbb{R}^4}dx\,A_\star^2(x)+\mbox{}\nonumber\\
  &\mbox{}-\frac1{48\pi^2}\,\Gamma(0,m^2/\Lambda^2)
  \int_{\mathbb{R}^4}dx\,A_\mu(x)\left\{\delta_{\mu\nu}\,\partial^2-\partial_\mu\partial_\nu\right\}A_\nu(x)
  +\ldots\,,
\end{align}
where the dots represent UV-finite contributions. Note that the first term exactly cancels the contribution of $\Gamma_{AA}^{(2)}$ (see eq.\ \eqref{div2nd}) so there are no quadratic divergences, as expected from gauge invariance.

Collecting all the divergences arising from $\Gamma_F^{(2)}$, $\Gamma_{AA}^{(2)}$ and $\Gamma_A^{(2)}$ we obtain
\begin{align}\label{divprop}
  \Gamma^{(2)}=
  \frac{11}{48\pi^2}\,\log{(\Lambda^2/m^2)}
  \int_{\mathbb{R}^4}dx\,A_\mu(x)\left\{\delta_{\mu\nu}\,\partial^2-\partial_\mu\partial_\nu\right\}A_\nu(x)
  +\ldots\,,
\end{align}
where the dots represent UV-finite terms.

\subsection{The $\beta$-function}\label{bf}

As shown in expression \eqref{divprop}, there are no quadratic UV divergences in the self-energy, so quantum fluctuations do not generate a mass term, which would break $U_\star(1)$ symmetry; instead, there is a logarithmic divergence which is removed by a charge renormalization \cite{Martin:1999aq}. Moreover, expression \eqref{divprop} is consistent with the transversality required by gauge symmetry.

The UV divergence in the quadratic part of the one-loop effective action can be removed by a redefinition of the physical coupling constant $e^2_R$ in the classical action~\eqref{theaction}, namely
\begin{align}
  \frac{1}{2e^2_R}=\frac{1}{2e^2}
  -\frac{11}{48\pi^2}\,\log{(\Lambda^2/m^2)}\,,
\end{align}
or, equivalently,
\begin{align}\label{er}
  e^2_R=e^2\left(1+\frac{11}{24\pi^2}\,e^2\,\log{(\Lambda^2/m^2)}\right)\,.
\end{align}
From this expression the $\beta$-function results
\begin{align}\label{beta}
  \beta(e):=\Lambda\,\partial_\Lambda e(\Lambda)=-\frac{11}{24\pi^2}\,e^3\,.
\end{align}
This expression, which shows that the theory is asymptotically free \cite{Krajewski:1999ja,SheikhJabbari:1999iw}, coincides with the $\beta$-function of pure Yang-Mills with a quadratic Casimir equal to $2$ in the adjoint representation.

At this point it is worth remarking an advantage of the background field method: as in ordinary (commutative) Yang-Mills theory, the $\beta$-function can be obtained directly from the divergences of the two-point function. In spite of this, to further illustrate the application of the master formula \eqref{masteraction}, we will verify in appendix \ref{3y4} that the same renormalization of the coupling constant is obtained either from the 3- or the 4-point functions.

\subsection{UV-finite part of the planar contributions to the self-energy}\label{pl}
The finite part of the photon self-energy receives contributions which have been omitted in the previous calculations (see eqs.~\eqref{div3rd} and~\eqref{div1st}) as well as contributions corresponding to non-planar diagrams, which will be considered in the next section.

Let us first consider the UV-finite contributions from eq.~\eqref{div3rd} to the planar part $\Gamma^{(2)}_F$,
\begin{align}
  &-\frac1{4\pi^2}
  \int d\bar\sigma\,\tilde F_{\mu\nu}(\sigma)\tilde F_{\mu\nu}(-\sigma)
  \int_{0}^\infty
  d\beta\,\frac{e^{-m^2\beta}}{\beta}\int_0^1 dt\,t
  \left(e^{-\beta\,t(1-t)\,\sigma^2}-1\right)\nonumber\\
  &=-\frac1{4\pi^2}
  \int d\bar\sigma\,\tilde F_{\mu\nu}(\sigma)\tilde F_{\mu\nu}(-\sigma)
  \sum_{n=1}^\infty
  (-1)^n\,\frac{\Gamma(n)\Gamma(n+2)}{\Gamma(2n+3)}
  \,\left(\frac{\sigma^2}{m^2}\right)^n\nonumber\\
  &=-\frac1{4\pi^2}
  \int d\bar\sigma\,\tilde F_{\mu\nu}(\sigma)\tilde F_{\mu\nu}(-\sigma)
  \left\{1-\sqrt{1+\frac{4m^2}{\sigma^2}}
  \ {\rm arcsinh}\sqrt{\frac{\sigma^2}{4m^2}}\right\}
  \,.
\end{align}
Note that we have removed the UV-cutoff by taking $\Lambda\rightarrow\infty$. If we consider the limit $m^2\rightarrow 0$ as well, we obtain
\begin{align}\label{fin3}
  -\frac1{4\pi^2}
  \int d\bar\sigma\,\tilde F_{\mu\nu}(\sigma)\tilde F_{\mu\nu}(-\sigma)
  \left\{1-\frac12\,\log{(\sigma^2/m^2)}+O(m^2/\sigma^2)\right\}\,.
\end{align}

Let us next consider the UV-finite contributions from eq.\ \eqref{div1st} to the planar contibution $\Gamma^{(2)}_{A}$,
\begin{align}
  &\frac1{2\pi^2}\int d\bar\sigma\,\tilde A_\mu(\sigma)\tilde A_\nu(-\sigma)
  \int_{0}^\infty
  \frac{d\beta}{\beta}\,e^{-m^2\beta}
  \int_0^1 dt_1\int_0^{t_1}dt_2\times\mbox{}\nonumber\\
  &\mbox{}\times\left\{-\frac{1}{2\beta}\,\delta_{\mu\nu}
  \,\sum_{n=2}^\infty\frac{(2\beta\,G_{12}\,\sigma^2)^n}{\Gamma(n+1)}
  -\dot G^2_{12}\,\sigma_\mu\sigma_\nu
  \,\sum_{n=1}^\infty\frac{(2\beta\,G_{12}\,\sigma^2)^n}{\Gamma(n+1)}\right\}\nonumber\\
  &=\frac1{24\pi^2}\int d\bar\sigma\,\tilde A_\mu(\sigma)\tilde A_\nu(-\sigma)
  \left(\delta_{\mu\nu}\,\sigma^2-\sigma_\mu\sigma_\nu\right)\times\mbox{}\nonumber\\
  &\mbox{}\times\left\{\frac{4m^2}{\sigma^2}+\frac43
  -\left(1+\frac{4m^2}{\sigma^2}\right)^{\frac32}{\rm arcsinh}\sqrt{\frac{\sigma^2}{4m^2}}\right\}
  \,.
\end{align}
In the limit $m^2\rightarrow 0$ this expression gives
\begin{align}\label{fin1}
  &\frac1{24\pi^2}\int d\bar\sigma\,\tilde A_\mu(\sigma)\tilde A_\nu(-\sigma)
  \left(\delta_{\mu\nu}\,\sigma^2-\sigma_\mu\sigma_\nu\right)\times\mbox{}\nonumber\\
  &\mbox{}\times\left\{\frac43-\frac12\,\log{(\sigma^2/m^2)}+O(m^2/\sigma^2)\right\}
  \,.
\end{align}
Altogether, expressions \eqref{fin3} and \eqref{fin1} give for the finite part of the planar contributions to $\Gamma^{(2)}$
\begin{align}
  \int d\bar\sigma\,\tilde A_\mu(\sigma)\tilde A_\nu(-\sigma)
  \left(\delta_{\mu\nu}\,\sigma^2-\sigma_\mu\sigma_\nu\right)
  \left\{\frac{11}{48\pi^2}\,\log{(\sigma^2/m^2)}-\frac4{9\pi^2}+O(m^2/\sigma^2)\right\}.
\end{align}
In consequence, there is a planar UV-finite but IR-divergent contribution to the effective action given by
\begin{align}\label{2pfinite}
  -\frac{11}{48\pi^2}\,\log{(\sigma^2/m^2)}
  \int_{\mathbb{R}^4}dx\,A_\mu(x)\left\{\delta_{\mu\nu}\,\partial^2-\partial_\mu\partial_\nu\right\}A_\nu(x)
  \,.
\end{align}
As is the case of ordinary comutative Yang-Mills theories, the dependence on the IR-cutoff $m^2$ can be cancelled by the corresponding term in eq.~\eqref{divprop}.

\subsection{Non-planar contributions to the photon self-energy}\label{npl}

To conclude our analysis of the photon self-energy, we study the non-planar contributions from the three types of quadratic terms in eq.\ \eqref{masteraction}, given by expressions \eqref{gamaf}, \eqref{gamaaa} and \eqref{gamaa}.

According to eq.\ \eqref{expval}, the non-planar part of expression \eqref{gamaf} is given by
\begin{align}
  \Gamma_{F}^{(2){\rm NP}}&=\frac{1}{4\pi^2}\int_{\Lambda^{-2}}^\infty
  \frac{d\beta}{\beta}\,e^{-m^2\beta}
  \int_0^1 dt\,t\times\mbox{}\nonumber\\
  &\mbox{}\times
  \int d\bar\sigma\,\tilde F_{\mu\nu}(\sigma)\tilde F_{\mu\nu}(-\sigma)
  \ e^{-\frac1{\beta}\,(\theta\sigma)^2-\beta\,t(1-t)\,\sigma^2}
  \,.
\end{align}
Integrating in $\beta$ we obtain
\begin{align}\label{np3rd}
  \Gamma_{F}^{(2){\rm NP}}=\frac1{2\pi^2}
  \int d\bar\sigma\,\tilde F_{\mu\nu}(\sigma)\tilde F_{\mu\nu}(-\sigma)
  \int_0^1 dt\,t\ K_0(2\sqrt{m^2+t(1-t)\sigma^2}\ |\theta\sigma|)
  \,.
\end{align}
Let us consider next the non-planar contribution from expression \eqref{gamaaa},
\begin{align}\label{np2nd}
  \Gamma_{AA}^{(2){\rm NP}}&=-\frac{1}{8\pi^2}\int_{0}^\infty
  d\beta\,\frac{e^{-m^2\beta}}{\beta^2}\int_0^1 dt
  \int d\bar\sigma\,\tilde A_\mu(\sigma)\tilde A_\mu(-\sigma)
  \ e^{-\frac1{\beta}\,(\theta\sigma)^2}\nonumber\\
  &=-\frac{1}{4\pi^2}
  \int d\bar\sigma\,\tilde A_\mu(\sigma)\tilde A_\mu(-\sigma)
  \ \frac{m}{|\theta\sigma|}\,K_1(2m|\theta\sigma|)\,.
\end{align}
This term would give a quadratic IR divergence with a tensorial structure $\delta_{\mu\nu}$. However, there is still one more contribution which completely cancels this term. In fact, using eq.\ \eqref{expvalp}, the non-planar contribution from expression \eqref{gamaa} reads
\begin{align}\label{np1st}
  &\Gamma_{A}^{(2){\rm NP}}=-\frac{1}{2\pi^2} \int_{\Lambda^{-2}}^\infty
  \frac{d\beta}{\beta}\,e^{-m^2\beta}
  \int_0^1 dt_1\int_0^{t_1}dt_2
  \int d\bar\sigma
  \,\tilde A_\mu(\sigma) \tilde A_\nu(-\sigma)\times\mbox{}\nonumber\\[2mm]
  &\mbox{}\times
  e^{-\frac1{\beta}\,(\theta\sigma)^2+2\beta\,G_{12}\,\sigma^2}
  \left\{-\frac1{2\beta}\,\delta_{\mu\nu}
  +\frac{1}{\beta^2}(\theta\sigma)_\mu(\theta\sigma)_\nu
  -\dot G^2_{12}\,\sigma_{\mu}\sigma_{\nu}\right\}\,.
\end{align}
After integrating in $\beta$ we obtain
\begin{align}
  &\Gamma_{A}^{(2){\rm NP}}=\frac1{2\pi^2}
  \int d\bar\sigma\,\tilde A_\mu(\sigma)\,\tilde A_\nu(-\sigma)
  \int_0^1 dt\,t\times\mbox{}\nonumber\\
  &\mbox{}\times
  \left\{\delta_{\mu\nu}\,\frac{\sqrt{m^2+t(1-t)\sigma^2}}{|\theta\sigma|}
  \,K_1\left(2\sqrt{m^2+t(1-t)\sigma^2}\,|\theta\sigma|\right)+\mbox{}\right.\nonumber\\
  &\mbox{}-2(\theta\sigma)_\mu(\theta\sigma)_\nu\left(\frac{m^2+t(1-t)\sigma^2}{(\theta\sigma)^2}\right)
  \,K_2\left(2\sqrt{m^2+t(1-t)\sigma^2}\,|\theta\sigma|\right)+\mbox{}\nonumber\\
  &\left.\mbox{}+\frac12\,\sigma_\mu\sigma_\nu\,\left(1-2t\right)^2
  \,K_0\left(2\sqrt{m^2+t(1-t)\sigma^2}\,|\theta\sigma|\right)\right\}
  \,.
\end{align}
Next, we use the identity \eqref{bi} derived in appendix \ref{bessels} to write this expression in the simplified form
\begin{align}\label{toget}
  &\Gamma_{A}^{(2){\rm NP}}=\frac1{4\pi^2}
  \int d\bar\sigma\,\tilde A_\mu(\sigma)\,\tilde A_\nu(-\sigma)
  \left\{\delta_{\mu\nu}\,\frac{m}{|\theta\sigma|}
  \,K_1\left(2m|\theta\sigma|\right)+\mbox{}\right.\nonumber\\
  &\mbox{}-\int_0^1 dt\,t
  \ \Big\{\left(1-2t\right)^2\left(\delta_{\mu\nu}\,\sigma^2-\sigma_\mu\sigma_\nu\right)
  \,K_0\left(2\sqrt{m^2+t(1-t)\sigma^2}\,|\theta\sigma|\right)+\mbox{}\nonumber\\
  &\left.\mbox{}+4(\theta\sigma)_\mu(\theta\sigma)_\nu\left(\frac{m^2+t(1-t)\sigma^2}{(\theta\sigma)^2}\right)
  \,K_2\left(2\sqrt{m^2+t(1-t)\sigma^2}\,|\theta\sigma|\right)\Big\}\right\}\,.
\end{align}
Note that, as already mentioned, the first term in this expression exactly cancels the whole contribution of $\Gamma_{AA}^{(2){\rm NP}}$, given by \eqref{np2nd}. The remaining terms, together with $\Gamma_{F}^{(2){\rm NP}}$ (cfr.\ eq.\ \eqref{np3rd}), give all non-planar contributions to the photon self-energy,
\begin{align}\label{npse}
  &\Gamma_{\rm {\rm NP}}^{(2)}=\frac1{\pi^2}
  \int d\bar\sigma\,\tilde A_\mu(\sigma)\,\tilde A_\nu(-\sigma)
  \int_0^1 dt\,t \times\mbox{}\nonumber\\
  &\mbox{}\times\Bigg\{\left(\delta_{\mu\nu}\,\sigma^2-\sigma_\mu\sigma_\nu\right)
  \left[1-\left(\tfrac12-t\right)^2\right]
  \,K_0\left(2\sqrt{m^2+t(1-t)\sigma^2}\,|\theta\sigma|\right)+\mbox{}\nonumber\\
  &\mbox{}-(\theta\sigma)_\mu(\theta\sigma)_\nu\left(\frac{m^2+t(1-t)\sigma^2}{(\theta\sigma)^2}\right)
  \,K_2\left(2\sqrt{m^2+t(1-t)\sigma^2}\,|\theta\sigma|\right)\Bigg\}\,.
\end{align}
Let us make some remarks regarding the content of this expression. Of course, the divergences of the Bessel functions for small values of $\sigma$ are a direct indication of the UV/IR mixing phenomenon and the consequent non-analyticity in $\theta$. However, the two terms in braces in expression \eqref{npse} present a quite different IR behaviour.

The term which has the tensorial structure $(\delta_{\mu\nu}\,\sigma^2-\sigma_\mu\sigma_\nu)$ shows a logarithmic IR divergence, even for $m^2\neq 0$, of the form
\begin{align}
  -\frac{11}{48\pi^2}
  \int d\bar\sigma\,\tilde A_\mu(\sigma)\,\tilde A_\nu(-\sigma)
  \left(\delta_{\mu\nu}\,\sigma^2-\sigma_\mu\sigma_\nu\right)
  \,\log{\left(m^2\,(\theta\sigma)^2\right)}+\ldots\,.
\end{align}
This result displays the correspondence between logarithmic IR divergences in non-planar contributions and UV divergences of the corresponding planar part \cite{Minwalla:1999px,Matusis:2000jf}. In addition, there are other tensorial structures in the non-planar part of the self-energy of the form $(\theta\sigma)_\mu(\theta\sigma)_\nu$ which, for small momentum $\sigma$, give \cite{Hayakawa:1999zf}
\begin{align}
  &-\frac1{4\pi^2}
  \int d\bar\sigma\,\tilde A_\mu(\sigma)\,\tilde A_\nu(-\sigma)
  \,\frac{(\theta\sigma)_\mu(\theta\sigma)_\nu}{(\theta\sigma)^4}+\ldots\,.
\end{align}
This quadratic IR divergence is not in correspondence with the UV divergence of the planar contributions \cite{Matusis:2000jf,Ruiz:2000hu}. In any case, all tensor structures in expression \eqref{npse} are transversal, in accordance with gauge symmetry. Note that this analysis can be carried out even for $m^2\neq 0$.

In the limit $m^2\rightarrow 0$, expression \eqref{npse} can be cast into the form
\begin{align}
  &\Gamma_{\rm {\rm NP}}^{(2)}=
  \int d\bar\sigma\,\tilde A_\mu(\sigma)\,\tilde A_\nu(-\sigma)
  \times\mbox{}\nonumber\\
  &\mbox{}\times\Bigg\{\left(\delta_{\mu\nu}\,\sigma^2-\sigma_\mu\sigma_\nu\right)
  \,\Sigma\left(|\sigma||\theta\sigma|\right)
  +\frac{(\theta\sigma)_\mu(\theta\sigma)_\nu}{(\theta\sigma)^4}
  \ \Xi\left(|\sigma||\theta\sigma|\right)\Bigg\}
  \,,
\end{align}
where
\begin{align}
  \Sigma(z)&=\frac{1}{8\pi^2\,z^3}\Big\{
  (4z^2+1)\left(\cosh{z}\ {\rm Shi}\,{z}-\sinh{z}\ {\rm Chi}\,{z}\right)+\mbox{}\nonumber\\
  &\mbox{}+z\left(\cosh{z}\ {\rm Chi}\,{z}-\sinh{z}\ {\rm Shi}\,{z}\right)-z\Big\}\,,
\end{align}
and
\begin{align}
  \Xi(z)&=-\frac{1}{8\pi^2\,z}\Big\{
  (z^2+3)\left(\cosh{z}\ {\rm Shi}\,{z}-\sinh{z}\ {\rm Chi}\,{z}\right)+\mbox{}\nonumber\\
  &\mbox{}+3z\left(\cosh{z}\ {\rm Chi}\,{z}-\sinh{z}\ {\rm Shi}\,{z}\right)-z\Big\}
  \,;
\end{align}
the functions ${\rm Shi},{\rm Chi}$ denote the hyperbolic sine and cosine integrals, respectively. Note that
\begin{align}
  \Sigma\left(|\sigma||\theta\sigma|\right)&=
  -\frac{11}{24\pi^2}\,\log{\left(|\sigma||\theta\sigma|\right)}+\ldots\,,\label{sigma}\\[2mm]
  \Xi\left(|\sigma||\theta\sigma|\right)&=
  -\frac{1}{4\pi^2}+\ldots\,,\label{xi}
\end{align}
as the external momentum $\sigma\rightarrow 0$. The coefficient $-\frac{11}{24\pi^2}$ in the logarithmic divergence \eqref{sigma} is related, as already mentioned, with the UV divergence of the planar contributions to the self-energy and thus provides the $\beta$-function of pure $U_\star(1)$. On the other hand, the negative sign in \eqref{xi} leads to the tachyonic instability originally described in \cite{Minwalla:1999px} and \cite{Matusis:2000jf}.

In conclusion, expression \eqref{npse} for the non-planar contributions to the self-energy shows that the polarization tensor is transversal, despite including a non-standard tensor structure $(\theta\sigma)_\mu(\theta\sigma)_\nu$, which is not Lorentz invariant. The photon propagator diverges at low energies, even when the IR-regulator is mantained, as a consequence of UV/IR mixing.

\section{Conclusions}\label{conclu}

We have applied the worldline formalism to pure noncommutative $U(1)$ gauge theory. The gauge invariance of the background field method yields a much more efficient computational tool in relation with the usual calculation of Feynman diagrams. In particular, the $\beta$-function---corresponding to the renormalization of the electric charge---can be computed directly from the (UV or IR) divergences of the photon self-energy. We have checked that the same charge renormalization is obtained from the 3- and 4-point functions. The result reproduces $U_\star(1)$ asymptotic freedom~\cite{Krajewski:1999ja,SheikhJabbari:1999iw}.

As an illustration of the efficiency of the method, we studied the two-point function, introducing both an UV- and an IR-cutoff, and we explicitly computed the photon self-energy for any value of the external momentum; the well-known behaviour for large and small momenta is reproduced. The polarization tensor is transversal although a non-standard tensorial structure, which is not present in Lorentz invariant models, arises due to non-planar contributions. Logarithmic IR divergences manifest the expected correspondence with UV singularities \cite{Minwalla:1999px,Matusis:2000jf}. However, there are also quadratic IR divergences, which do not have an UV counterpart and lead to the tachyonic instability described in \cite{Minwalla:1999px,Matusis:2000jf}. All results were obtained without removing the IR-cutoff $m^2$.

Concerning the implementation of the worldline formalism in this nonlocal gauge theory, there are two technical issues that are worth mentioning. The use of phase space path integrals to determine spectral quantities of nonlocal operators appears promising as long as one can overcome the difficulties originated in the ordering ambiguities of noncommuting operators. Of course, this also happens in local theories: the computation of the corresponding counterterms is an unavoidable task in the application of the worldline formalism to quantum fields on curved spacetimes~\cite{Bastianelli:2006rx}. We have shown that the nonlocal operators relevant in noncommutative gauge theories are already Weyl-ordered so, remarkably, no counterterms have to be introduced if one appropriately expresses all operators in terms of Moyal products. The second point we would like to address is the independence of the trace computation on the worldline Green's function (proved in appendix \ref{pbc}); this allowed us to exploit translation invariance (in the worldline proper time) to notably simplify all calculations.

The present study of a noncommutative gauge theory from the worldline perspective has proved very efficient in the computation of the effective action. Our research program goes on into the study of noncommutative gauge fields in the Grosse-Wulkenhaar context with worldline techniques.

Finally, it would also be quite interesting to apply the present worldline approach to the study of noncommutative extensions of Einstein gravity~\cite{Chamseddine:1992yx, Aschieri:2005yw}. In fact the approach described in the present manuscript was already used to study commutative perturbative quantum gravity~\cite{Bastianelli:2013tsa}.

\section*{Acknowledgments}{The authors thank Idrish Huet and Christian Schubert for helpful discussions, and acknowledge the partial support of the grants ``Fondo Institucional de CONACYT (FOINS) n.\ 219773'' and ``Programa de Cooperación Bilateral MINCyT -- CONACYT (ME/13/16)''. N.A.\ was supported by the PROMEP grant DSA/ 103.5/14/11184. O.C.\ was partly supported by the UCMEXUS–CONACYT grant CN-12-564}. D.D.\ and P.P.\ thank support from CONICET, Argentina; financial support from
CONICET (PIP 0681), ANPCyT (PICT 0605) and UNLP (X615) is also acknowledged.

\appendix

\section{Worldline Green's functions}\label{pbc}

In this section we compare the mean value
\begin{align}\label{mv}
  \left\langle\,e^{i\int_0^1 dt\left\{k(t)p(t)+j(t)x(t)\right\}}\,\right\rangle
  =\frac{ \int\mathcal{D}x(t)\mathcal{D}p(t)\ e^{-\int_0^1 dt\,\left\{p^2-ip\dot{x}\right\}}
  \,e^{i\int_0^1 dt\left\{kp+jx\right\}}}
  {\int\mathcal{D}x(t)\mathcal{D}p(t)\ e^{-\int_0^1 dt\,\left\{p^2-ip\dot{x}\right\}}}\,,
\end{align}
as computed for two types of conditions on the phase space trajectories $x(t),p(t)$. Firstly, we will study trajectories with homogeneous Dirichlet conditions in configuration space, $x(0)=x(1)=0$. Secondly,  we will impose periodic boundary conditions on both $x(t)$ and $p(t)$; however, since these conditions involve a zero mode, we will integrate over the subspace of trajectories which are orthogonal to this zero mode.

Let us first write expression \eqref{mv} as
\begin{align}\label{mvm}
  \left\langle\,e^{i\int_0^1 dt\left\{k(t)p(t)+j(t)x(t)\right\}}\,\right\rangle
  =\frac{\int\mathcal{D}Z(t)\ e^{-\frac12\int_0^1 dt\,Z(t)^T\,D\,Z(t)}\,e^{i\int_0^1 dt\,Z(t)^T J(t)}}
  {\int\mathcal{D}Z(t)\ e^{-\int_0^1 dt\,Z(t)^T\,D\,Z(t)}}\,,
\end{align}
where $Z(t)$ denote trajectories in phase space and $J(t)$ the corresponding external sources,
\begin{align}
  Z(t)=\left(\begin{array}{c}p(t)\\x(t)\end{array}\right)\,,\qquad
  J(t)=\left(\begin{array}{c}k(t)\\j(t)\end{array}\right)\,.
\end{align}
We have also defined the matricial operator
\begin{align}
  D=\left(\begin{array}{cc}2&-i\partial_t\\[2mm]
  i\partial_t&0\end{array}\right)\,.
\end{align}
Under Dirichlet boundary conditions $D$ is symmetric and invertible. Its inverse is given by
\begin{align}
  D_{\rm Dir}^{-1}=\left(\begin{array}{ccc}
  \frac12
  &&-\frac{i}{2}\epsilon(t-t')-it'+\frac{i}{2}\\[2mm]
  \frac{i}{2}\epsilon(t-t')-it+\frac{i}{2}
  &&-|t-t'|-2tt'+t+t'
  \end{array}\right)\,.
\end{align}
On the other hand, for string-inspired (periodic) boundary conditions $D$ is also symmetric but has a zero mode,
\begin{align}
  Z_0(t)=\left(\begin{array}{c}0\\1\end{array}\right)\,,
\end{align}
and is therefore not invertible. However, the inverse in the subspace orthogonal to $Z_0(t)$ is given by
\begin{align}
  D_{\rm per}^{-1}=\left(\begin{array}{ccc}
  \frac12
  &&-\frac{i}{2}\epsilon(t-t')+i(t-t')\\[2mm]
  \frac{i}{2}\epsilon(t-t')-i(t-t')
  &&-|t-t'|-2tt'+t^2+t'^2+\frac16
  \end{array}\right)\,.
\end{align}
We can now complete squares in expression \eqref{mvm} and express the mean value in terms of the inverse operators $D^{-1}$. The result reads
\begin{align}\label{meanvalueapp}
  &\left\langle\,e^{i\int_0^1 dt\left\{k(t)p(t)+j(t)x(t)\right\}}\,\right\rangle
  =e^{-\frac12\int_0^1 dt\,J(t)^T\,D^{-1}\,J(t)}
  \nonumber\\
  &=\exp{\left(-\int\int dtdt'\left\{\tfrac14\,k(t)k(t')+g(t,t')j(t)j(t')
  +\tfrac{i}{2}\, h(t,t')k(t)j(t')\right\}\right)}
  \,,
\end{align}
with
\begin{align}
  g(t,t')&:=-\frac12|t-t'|-tt'+\frac12t+\frac12t'\,,\\ h(t,t')&:=2\partial_t g(t,t')=-\epsilon(t-t')-2t'+1\,,
\end{align}
for Dirichlet boundary conditions and
\begin{align}
  g_{\rm per}(t,t')&:=-\frac12|t-t'|-tt'+\frac12t^2+\frac12t'^2+\frac1{12}\,,\\
  h_{\rm per}(t,t')&:=2\partial_t g_{\rm per}(t,t') = -\epsilon(t-t')+2(t-t')\,,
\end{align}
for string-inspired boundary conditions. Note that if the mean value \eqref{meanvalueapp} is used to compute the heat-trace then both for Dirichlet and for string-inspired boundary conditions the external current satisfies
\begin{align}\label{q=0}
    \int_0^1 dt\,j(t)=0\,.
\end{align}
In the case of Dirichlet conditions such restriction results from the fact that the heat-trace involves an integration over the configuration space variable $x$ that enforces~\eqref{q=0}. For string-inspired boundary conditions the same constraint arises from the integration over the zero mode. In both cases, in Fourier space, condition~\eqref{q=0} corresponds to the conservation of the total four-momentum. It is clear that under condition~\eqref{q=0}, the mean value~\eqref{meanvalueapp} does not depend on the chosen boundary conditions. On the other hand, if one computes local quantities the difference between both types of boundary conditions arises in total derivative terms.

Finally, since under condition~\eqref{q=0} terms in $g(t,t')$ which depend only on $t$ or on $t'$ (but not on both), as well as terms in $h(t,t')$ which do not depend on $t'$, are irrelevant in~\eqref{meanvalueapp}, we can instead use the simplified Green's functions:
\begin{align}
  G(t-t')&:=-\frac12|t-t'|+\frac12(t-t')^2\,,\\
  H(t-t')&:= 2\dot G(t-t')=-\epsilon(t-t')+2(t-t')\,,
\end{align}
that are homogeneous string-inspired Green's functions.

\section{3- and 4-point functions}\label{3y4}

In this appendix we compute the remaining UV divergences of the one-loop effective action which, as shown by expression \eqref{masteraction}, are due exclusively to planar contributions to the 3- and 4-point functions.

Only three types of cubic terms in expression \eqref{masteraction} give UV-divergent contributions: terms of the form $(V^A)^3$, of the form $V^A\,V^{AA}$, and the term $(V^F)^2$, already computed in \eqref{FFdiv}. The first of these terms gives the following contribution:
\begin{align}
  &\frac{1}{16\pi^2}\int_{\Lambda^{-2}}^\infty
  d\beta\,e^{-m^2\beta}
  \int_0^1 dt_1\int_0^{t_1}dt_2\int_0^{t_{2}}dt_3
  \,\times\mbox{}\nonumber\\[2mm]
  &\mbox{}\times
  \int_{\mathbb{R}^4} dx\left\langle
  \left(\frac{2}{\sqrt{\beta}}\,p_\mu(t_1)\,V_\mu^A(t_1)
  \frac{2}{\sqrt{\beta}}\,p_\mu(t_2)\,V_\mu^A(t_2)
  \frac{2}{\sqrt{\beta}}\,p_\mu(t_3)\,V_\mu^A(t_3)\right)\right\rangle\nonumber\\[2mm]
  &=\frac{i}{2\pi^2}\int_{\Lambda^{-2}}^\infty
  d\beta\,e^{-m^2\beta}
  \int_0^1 dt_1\int_0^{t_1}dt_2\int_0^{t_{2}}dt_3
  \int d\bar\sigma_1d\bar\sigma_2d\bar\sigma_3
  \,\times\mbox{}\nonumber\\[2mm]
  &\mbox{}\times
  \,\bar\delta(\sigma_1+\sigma_2+\sigma_3)\,
  \tilde A_\mu(\sigma_1)\tilde A_\nu(\sigma_2)\tilde A_\tau(\sigma_3)
  \ e^{-\beta\sum_{i,j}G_{ij}\,\sigma_i\sigma_j}
  \,\times\mbox{}\nonumber\\[2mm]
  &\mbox{}\times
  \,\frac{\partial}{\partial\rho_{1\mu}}\,\frac{\partial}{\partial\rho_{2\nu}}\,\frac{\partial}{\partial\rho_{3\tau}}
  \left.e^{-\sum_{i,j} \left\{\frac1{4\beta}\,\rho_i\rho_j+i\dot G_{ij}\,\rho_i\sigma_j\right\}}
  \right|_{\rho_i=-\theta\sigma_i}-(\theta\rightarrow-\theta)
  \,.
\end{align}
As already discussed, the only effect of the Bern-Kosower form factor is to implement the (time-ordered) $\star$-product of the fields, apart from the $\theta$-independent term, which is $O(\beta^0)$. The three derivatives give a leading contribution (for small $\beta$) of the form $\frac{i}{2\beta}\delta_{\mu\nu}\dot G_{3j}\sigma_{j\tau}$, together with the corresponding permutations. The divergent contribution then reads
\begin{align}\label{vavava}
  &-\frac{1}{4\pi^2}\,\log{(\Lambda^2/m^2)}
  \int d\bar\sigma_1d\bar\sigma_2d\bar\sigma_3
  \,\bar\delta(\sigma_1+\sigma_2+\sigma_3)
  \,\times\mbox{}\nonumber\\[2mm]
  &\mbox{}\times
  \,\tilde A_\mu(\sigma_1)\tilde A_\nu(\sigma_2)\tilde A_\tau(\sigma_3)
  \left\{e^{i\,\sum_{i<j} \sigma_i\theta\sigma_j}-e^{-i\,\sum_{i<j} \sigma_i\theta\sigma_j}\right\}
  \,\times\mbox{}\nonumber\\[2mm]
  &\mbox{}\times
  \int_0^1 dt_1\int_0^{t_1}dt_2\int_0^{t_{2}}dt_3
  \left\{\delta_{\mu\nu}\,\dot G_{3j}\sigma_{j\tau}
  +\delta_{\mu\tau}\,\dot G_{2j}\sigma_{j\nu}
  +\delta_{\nu\tau}\,\dot G_{1j}\sigma_{j\mu}\right\}\nonumber\\[2mm]
  &=-\frac{1}{24\pi^2}\,\log{(\Lambda^2/m^2)}
  \int d\bar\sigma_1d\bar\sigma_2d\bar\sigma_3
  \,\bar\delta(\sigma_1+\sigma_2+\sigma_3)
  \,\times\mbox{}\nonumber\\[2mm]
  &\mbox{}\times
  \,\tilde A_\mu(\sigma_1)\tilde A_\nu(\sigma_2)\tilde A_\nu(\sigma_3)\,\sigma_{3\mu}
  \left\{e^{i\,\sum_{i<j} \sigma_i\theta\sigma_j}-e^{-i\,\sum_{i<j} \sigma_i\theta\sigma_j}\right\}
  \,.
\end{align}
The second type of divergent contribution to the three-point function is given by
\begin{align}
  &-\frac{1}{16\pi^2}\int_{\Lambda^{-2}}^\infty
  \frac{d\beta}{\beta}\,e^{-m^2\beta}
  \int_0^1 dt_1\int_0^{t_1}dt_2
  \,\times\mbox{}\nonumber\\[2mm]
  &\mbox{}\times
  \int_{\mathbb{R}^4} dx\left\langle
  \frac{2}{\sqrt{\beta}}\,p_\mu(t_1)\,V^{A}_\mu(t_1)\,V^{AA}(t_2)+\left(t_1\leftrightarrow t_2\right)\right\rangle\,.
\end{align}
However, the planar part of this expression represents the difference between the mean values $\langle p_\mu A_\mu \,A^2_\star\rangle$ for both signs of $\theta$. But, as we have seen, the only effect of $\theta$ in this planar mean value is to provide the $\star$-product between $A_\mu$ and $A^2_\star$ which, under the integral sign, can be removed. Therefore, the difference between the term with $\theta$ and the term with $-\theta$ vanishes.

In consequence, the contribution computed in \eqref{vavava} together with the cubic part of \eqref{FFdiv} gives the full UV divergence of the three-point function,
\begin{align}
  &\frac{11}{24\pi^2}\,\log{(\Lambda^2/m^2)}
  \int d\bar\sigma_1d\bar\sigma_2d\bar\sigma_3
  \,\bar\delta(\sigma_1+\sigma_2+\sigma_3)
  \,\times\mbox{}\nonumber\\[2mm]
  &\mbox{}\times
  \,\tilde A_\mu(\sigma_1)\tilde A_\nu(\sigma_2)\tilde A_\nu(\sigma_3)\,\sigma_{3\mu}
  \left(e^{i\,\sum_{i<j} \sigma_i\theta\sigma_j}-e^{-i\,\sum_{i<j} \sigma_i\theta\sigma_j}\right)\,.
\end{align}
This logarithmically divergent contribution can be absorbed in the cubic term of expression \eqref{theactioninF} by the same infinite redefinition of the coupling constant $e_R^2$ given in \eqref{er} and derived from the divergences of the photon self-energy.

We finally check that the same charge renormalization is obtained from the studies of the divergences of the four-point function. The terms that contribute to these divergences are of the form $(V^A)^4$, $(V^A)^2V^{AA}$, $(V^{AA})^2$, and $(V^F)^2$. The latter was already computed in~\eqref{FFdiv}. Let us thus compute the other three contributions. The first one reads
\begin{align}
  &-\frac{1}{16\pi^2}\int_{\Lambda^{-2}}^\infty
  d\beta\,\beta\, e^{-m^2\beta}
  \int_0^1 dt_1\ldots\int_0^{t_{3}}dt_4
  \,\times\mbox{}\nonumber\\[2mm]
  &\mbox{}\times
  \int_{\mathbb{R}^4} dx\left\langle
  \frac{2}{\sqrt{\beta}}\,p_\mu(t_1)\,V_\mu^A(t_1)\ldots \frac{2}{\sqrt{\beta}}\,p_\mu(t_4)\,V_\mu^A(t_4)
  \right\rangle\nonumber\\[2mm]
  &=-\frac{1}{\pi^2}\int_{\Lambda^{-2}}^\infty
  d\beta\,\beta\, e^{-m^2\beta}
  \int_0^1 dt_1\ldots\int_0^{t_{3}}dt_4
  \int d\bar\sigma_1\ldots d\bar\sigma_4
  \,\times\mbox{}\nonumber\\[2mm]
  &\mbox{}\times
  \,\bar\delta(\sigma_1+\ldots+\sigma_4)
  \,\tilde A_\mu(\sigma_1)\tilde A_\nu(\sigma_2)\tilde A_\tau(\sigma_3)\tilde A_\omega(\sigma_4)
  \ e^{-\beta\sum_{i,j}G_{ij}\,\sigma_i\sigma_j}
  \times\mbox{}\nonumber\\[2mm]
  &\mbox{}\times
  \frac{\partial}{\partial\rho_{1\mu}}\ldots
  \frac{\partial}{\partial\rho_{4\omega}}
  \left.e^{-\sum_{i,j} \left\{\frac1{4\beta}\,\rho_i\rho_j+i\dot G_{ij}\,\rho_i\sigma_j\right\}}
  \right|_{\rho_i=-\theta\sigma_i}+\left(\theta\rightarrow -\theta\right)  \,.
\end{align}
After performing the four derivatives, the leading contribution (for small $\beta$) is of the form $(-\frac{1}{2\beta})^2\,\delta_{\mu\nu}\delta_{\tau\omega}$, together with the corresponding permutations. Therefore, the divergent part of this contribution reads
\begin{align}\label{4paaaa}
  &-\frac{1}{96\pi^2}\,\log{(\Lambda^2/m^2)}
  \int d\bar\sigma_1\ldots d\bar\sigma_4
  \,\bar\delta(\sigma_1+\ldots+\sigma_4)
  \,\times\mbox{}\nonumber\\[2mm]
  &\mbox{}\times
  \,\tilde A_\mu(\sigma_1)\tilde A_\nu(\sigma_2)\tilde A_\tau(\sigma_3)\tilde A_\omega(\sigma_4)
  \left(e^{i\,\sum_{i<j} \sigma_i\theta\sigma_j}+e^{-i\,\sum_{i<j} \sigma_i\theta\sigma_j}\right)
  \times\mbox{}\nonumber\\[2mm]
  &\mbox{}\times\left\{
  \delta_{\mu\nu}\delta_{\tau\omega}+\delta_{\mu\tau}\delta_{\nu\omega}+\delta_{\mu\omega}\delta_{\nu\tau}\right\}
  \nonumber\\[2mm]
  &=-\frac{1}{48\pi^2}\,\log{(\Lambda^2/m^2)}
  \int d\bar\sigma_1\ldots d\bar\sigma_4
  \,\bar\delta(\sigma_1+\ldots+\sigma_4)
  \ e^{i\,\sum_{i<j} \sigma_i\theta\sigma_j}
  \,\times\mbox{}\nonumber\\[2mm]
  &\mbox{}\times
  \left(2\,\tilde A_\mu(\sigma_1)\tilde A_\mu(\sigma_2)\tilde A_\nu(\sigma_3)\tilde A_\nu(\sigma_4)
  +\tilde A_\mu(\sigma_1)\tilde A_\nu(\sigma_2)\tilde A_\mu(\sigma_3)\tilde A_\nu(\sigma_4)\right)\,.
\end{align}
The contribution of the term of the form $(V^A)^2V^{AA}$ is given by
\begin{align}
  &\frac{1}{16\pi^2}\int_{\Lambda^{-2}}^\infty
  d\beta\,e^{-m^2\beta}
  \int_0^1 dt_1\int_0^{t_1}dt_2 \int_0^{t_2}dt_3
  \int_{\mathbb{R}^4} dx
  \times\mbox{}\nonumber\\[2mm]
  &\mbox{}\times
  \ \left\langle
  \frac{2}{\sqrt{\beta}}\,p_\mu(t_1)\,V_\mu^A(t_1)\ \frac{2}{\sqrt{\beta}}\,p_\mu(t_2)\,V_\mu^A(t_2)\ V^{AA}(t_3)
  +\left(t_2\leftrightarrow t_3\right)+\left(t_1\leftrightarrow t_3\right)
  \right\rangle\nonumber\\[2mm]
  &=-\frac{1}{4\pi^2}\int_{\Lambda^{-2}}^\infty
  d\beta\,e^{-m^2\beta}
  \int_0^1 dt_1\int_0^{t_1}dt_2 \int_0^{t_2}dt_3
  \,\times\mbox{}\nonumber\\[2mm]
  &\mbox{}\times
  \int d\bar\sigma_1 d\bar\sigma_2 d\bar\sigma_3\,\bar\delta(\sigma_1+\sigma_2+\sigma_3)
  \,\tilde A_\mu(\sigma_1)\tilde A_\nu(\sigma_2)\tilde A^2_\star(\sigma_3)
  \ e^{-\beta\sum_{i,j}G_{ij}\,\sigma_i\sigma_j}
  \times\mbox{}\nonumber\\[2mm]
  &\mbox{}\times
  \ \left(\frac{\partial}{\partial\rho_{1\mu}}\frac{\partial}{\partial\rho_{2\nu}}
  \left.e^{-\sum_{i,j} \left\{\frac1{4\beta}\,\rho_i\rho_j+i\dot G_{ij}\,\rho_i\sigma_j\right\}}
  \right|_{\rho_i=-\theta\sigma_i}
  +\left(\sigma_2\leftrightarrow \sigma_3\right)+\left(\sigma_1\leftrightarrow \sigma_3\right)\right)
  +\mbox{}\nonumber\\[2mm]
  &\mbox{}+\left(\theta\rightarrow-\theta\right)\,.
\end{align}
After performing the derivatives, the leading contribution is given by $-\frac{1}{2\beta}\delta_{\mu\nu}$ so the divergent part of this contribution reads
\begin{align}\label{4paaa2}
  &\frac{1}{8\pi^2}\,\log{(\Lambda^2/m^2)}
  \int_0^1 dt_1\int_0^{t_1}dt_2 \int_0^{t_2}dt_3
  \,\times\mbox{}\nonumber\\[2mm]
  &\mbox{}\times
  \int d\bar\sigma_1 d\bar\sigma_2 d\bar\sigma_3\,\bar\delta(\sigma_1+\sigma_2+\sigma_3)
  \,\tilde A_\mu(\sigma_1)\tilde A_\mu(\sigma_2)\tilde A^2_\star(\sigma_3)
  \times\mbox{}\nonumber\\[2mm]
  &\mbox{}\times
  \ \left(e^{i\,\sum_{i<j} \sigma_i\theta\sigma_j}+e^{-i\,\sum_{i<j} \sigma_i\theta\sigma_j}
  +\left(\sigma_2\leftrightarrow \sigma_3\right)+\left(\sigma_1\leftrightarrow \sigma_3\right)\right)
  \nonumber\\[2mm]
  &=\frac{1}{8\pi^2}\,\log{(\Lambda^2/m^2)}
  \int d\bar\sigma_1 d\bar\sigma_2 d\bar\sigma_3\,\bar\delta(\sigma_1+\sigma_2+\sigma_3)
  \,\times\mbox{}\nonumber\\[2mm]
  &\mbox{}\times
  \,\tilde A_\mu(\sigma_1)\tilde A_\mu(\sigma_2)\tilde A^2_\star(\sigma_3)
  \ e^{i\,\sum_{i<j} \sigma_i\theta\sigma_j}\,.
\end{align}
Finally the contribution of the $(V^{AA})^2$ term is given by
\begin{align}
  &-\frac{1}{16\pi^2}\int_{\Lambda^{-2}}^\infty
  \frac{d\beta}{\beta}\,e^{-m^2\beta}
  \int_0^1 dt_1\int_0^{t_1}dt_2
  \int_{\mathbb{R}^4} dx\left\langle\left(V^{AA}(t_1)\ V^{AA}(t_2)\right)\right\rangle\nonumber\\[2mm]
  &=-\frac{1}{8\pi^2}\int_{\Lambda^{-2}}^\infty
  \frac{d\beta}{\beta}\,e^{-m^2\beta}
  \int_0^1 dt_1\int_0^{t_1}dt_2
  \int d\bar\sigma\,\tilde A^2_\star(\sigma)\tilde A^2_\star(-\sigma)
  e^{-\beta\sum_{i,j}G_{ij}\,\sigma_i\sigma_j}\,,
\end{align}
whose divergent part reads
\begin{align}\label{4pa2a2}
  &-\frac{1}{16\pi^2}\,\log{(\Lambda^2/m^2)}
  \int d\bar\sigma\,\tilde A^2_\star(\sigma)\tilde A^2_\star(-\sigma)\,.
\end{align}
Collecting the results of eqs.\ \eqref{4paaaa}, \eqref{4paaa2}, \eqref{4pa2a2} and \eqref{FFdiv} we obtain for the UV divergence of the 4-point function
\begin{align}
  &-\frac{11}{48\pi^2}\,\log{(\Lambda^2/m^2)}
  \int d\bar\sigma_1\ldots d\bar\sigma_4
  \,\bar\delta(\sigma_1+\ldots+\sigma_4)
  \ e^{i\,\sum_{i<j} \sigma_i\theta\sigma_j}
  \,\times\mbox{}\nonumber\\[2mm]
  &\mbox{}\times
  \left\{\tilde A_\mu(\sigma_1)\tilde A_\mu(\sigma_2)\tilde A_\nu(\sigma_3)\tilde A_\nu(\sigma_4)
  -\tilde A_\mu(\sigma_1)\tilde A_\nu(\sigma_2)\tilde A_\mu(\sigma_3)\tilde A_\nu(\sigma_4)\right\}\,,
\end{align}
which, as expected, is cancelled by the renormalization of the charge given by expression~\eqref{er}.

\section{Bessel functions}\label{bessels}

In this last appendix we prove an identity which we employed to get eq.\ \eqref{toget} and show that non-planar contributions to the self-energy are transversal. We begin by noting that
\begin{align}
  &\partial_t\left\{t(1-2t)\sqrt{m^2+t(1-t)\sigma^2}
  \,K_1\left(2|\theta\sigma|\,\sqrt{m^2+t(1-t)\sigma^2}\right)\right\}=\nonumber\\
  &=-|\theta\sigma|\,\sigma^2\,t(1-2t)^2
  \,K_0\left(2|\theta\sigma|\,\sqrt{m^2+t(1-t)\sigma^2}\right)+\mbox{}\nonumber\\
  &\mbox{}+\left\{(1-2t)-2t\right\}\sqrt{m^2+t(1-t)\sigma^2}
  \,K_1\left(2|\theta\sigma|\,\sqrt{m^2+t(1-t)\sigma^2}\right)\,.
\end{align}
Since the first term in braces in the last line of this expression---proportional to $(1-2t)$---is odd under the interchange $t\leftrightarrow 1-t$, then its integral in the interval $t\in[0,1]$ vanishes; the integral of the remaining terms give the mentioned identity,
\begin{align}\label{bi}
  \frac{m}{2|\theta\sigma|}\,K_1\left(2m|\theta\sigma|\right)&=
  \int_0^1 dt\,t\left\{\frac12\,\sigma^2\,(1-2t)^2
  \,K_0\left(2|\theta\sigma|\,\sqrt{m^2+t(1-t)\sigma^2}\right)+\mbox{}\right.\nonumber\\
  &\left.\mbox{}+\frac{\sqrt{m^2+t(1-t)\sigma^2}}{|\theta\sigma|}
  \,K_1\left(2|\theta\sigma|\,\sqrt{m^2+t(1-t)\sigma^2}\right)\right\}\,.
\end{align}


\end{document}